\documentclass[aps,prx,reprint,twocolumn,superscriptaddress,floatfix,nofootinbib,longbibliography]{revtex4-2}
\usepackage{graphicx,amsmath,amsfonts,amssymb,amsthm,xr}

\usepackage{epsfig,amsmath,amssymb,color,dsfont,upgreek,physics}
\usepackage{mathrsfs}
\usepackage{mathtools}
\usepackage{bbold}
\usepackage{comment}

\usepackage{float}
\usepackage[caption=false]{subfig}

\usepackage[dvipsnames]{xcolor}
\usepackage[bookmarks=true,colorlinks,linkcolor=OrangeRed,urlcolor=NavyBlue,citecolor=RoyalBlue]{hyperref}
\usepackage{orcidlink}
\usepackage{quantikz}
\usepackage{helvet}
\usepackage[tracking=true]{microtype}
\SetTracking{encoding=*}{100} 

\definecolor{limeGreen}{rgb}{0.55, 0.71, 0.0}
\definecolor{mygold}{rgb}{0.5,0.6,0.7}

\definecolor{mypurple}{rgb}{0.49,0.18,0.56}

\begin{document}

\title{Probing Hadron Scattering in Lattice Gauge Theories on Qudit Quantum Computers}
\author{Rohan Joshi}
\affiliation{Max Planck Institute of Quantum Optics, 85748 Garching, Germany}
\affiliation{Munich Center for Quantum Science and Technology (MCQST), 80799 Munich, Germany}

\author{Jan C.~Louw${}^{\orcidlink{0000-0002-5111-840X}}$}
\affiliation{Max Planck Institute of Quantum Optics, 85748 Garching, Germany}
\affiliation{Munich Center for Quantum Science and Technology (MCQST), 80799 Munich, Germany}

\author{Michael Meth${}^{\orcidlink{0000-0002-5446-3962}}$}
\affiliation{Universit\"at Innsbruck, Institut f\"ur Experimentalphysik, 6020 Innsbruck, Austria}

\author{Jesse J.~Osborne${}^{\orcidlink{0000-0003-0415-0690}}$}
\affiliation{Max Planck Institute of Quantum Optics, 85748 Garching, Germany}
\affiliation{Munich Center for Quantum Science and Technology (MCQST), 80799 Munich, Germany}

\author{Kevin Mato}
\affiliation{Chair for Design Automation, Technical University of Munich, Munich, Germany}

\author{Guo-Xian Su${}^{\orcidlink{0000-0001-7936-762X}}$}
\affiliation{Department of Physics, Massachusetts Institute of Technology, Cambridge, MA 02139, USA}
\affiliation{MIT-Harvard Center for Ultracold Atoms, Cambridge, MA 02139, USA}

\author{Martin Ringbauer${}^{\orcidlink{0000-0001-5055-6240}}$}
\email{martin.ringbauer@uibk.ac.at}
\affiliation{Universit\"at Innsbruck, Institut f\"ur Experimentalphysik, 6020 Innsbruck, Austria}

\author{Jad C.~Halimeh${}^{\orcidlink{0000-0002-0659-7990}}$}
\email{jad.halimeh@physik.lmu.de}
\affiliation{Max Planck Institute of Quantum Optics, 85748 Garching, Germany}
\affiliation{Department of Physics and Arnold Sommerfeld Center for Theoretical Physics (ASC), Ludwig Maximilian University of Munich, 80333 Munich, Germany}
\affiliation{Munich Center for Quantum Science and Technology (MCQST), 80799 Munich, Germany}
\begin{abstract}
An overarching goal in the flourishing field of quantum simulation for high-energy physics is the first-principles study of the microscopic dynamics of scattering processes on a quantum computer. Currently, this is hampered by small system sizes and a restriction to two-level representations of the gauge fields in state-of-the-art quantum simulators. Here, we propose efficient experimentally feasible digital qudit quantum circuits for far-from-equilibrium quench dynamics of a $\mathrm{U}(1)$ quantum link lattice gauge theory, where the electric and gauge fields are represented as spin-$1$ operators. Using dedicated numerical simulations, we probe scattering processes in this model on these proposed circuits, focusing on meson-meson and meson-antimeson collisions. The latter are not possible with a two-level representation of the fields, highlighting the suitability of qudits in exploring scattering processes relevant to quantum electrodynamics. The probed scattering dynamics showcases rich physics, including meson flipping and a reflection-transmission transition in meson-antimeson collisions as a function of the gauge coupling strength. Our simulations, which include realistic noise models of dephasing and depolarization, show very good agreement with the exact noiseless dynamics, signaling the readiness of current qudit platforms to observe microscopic scattering dynamics with significantly shallower circuit depths than their qubit counterparts.
\end{abstract}

\maketitle

\tableofcontents

\section{Introduction}
\label{Into}
Scattering experiments are central to the study of high-energy physics (HEP)~\cite{Weinberg_book,Zee_book,Peskin2016}. They facilitate the breakdown of matter into its most fundamental constituents, including quarks and gluons, allowing the study of their subsequent interactions and hadronization~\cite{Ellis_book}. Dedicated particle colliders such as the Large Hadron Collider (LHC) at CERN and the Relativistic Heavy Ion Collider (RHIC) at Brookhaven National Laboratory perform scattering experiments that provide a glimpse of the inner workings of nature. The collected data contributes critically to our knowledge of subatomic structures, the discovery of new particles~\cite{ATLAS2012,CMS2012}, and our understanding of the evolution of the early universe by creating a quark-gluon plasma at high-energy ion collisions~\cite{Adcox2005,Back2005,Arsene2005}.

\begin{figure*}[t]
\captionsetup[subfloat]{position=top,justification=raggedright,singlelinecheck=false}
    \centering
    \captionsetup[subfloat]{oneside,margin={0cm,0cm}}
    \subfloat[\scriptsize \sffamily\bfseries\textls{Probing scattering on a qudit processor}]{%
        \includegraphics[width=\textwidth]{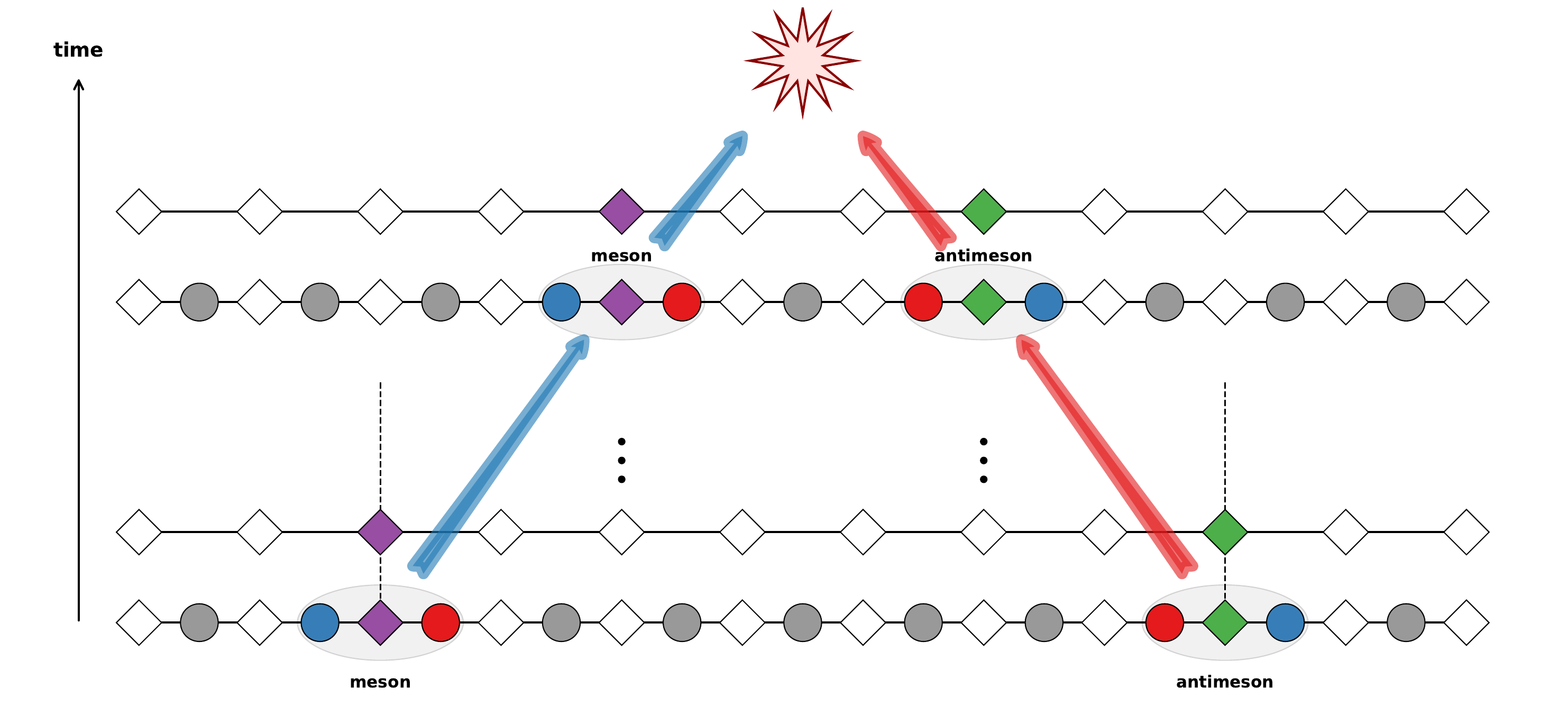}
    }
    \hfill
\captionsetup[subfloat]{oneside,margin={-0.1cm,0cm},skip=0.5cm}
\subfloat[\scriptsize\sffamily\bfseries\textls{Gauge-invariant states}]{%
       \includegraphics[width=0.51\textwidth]{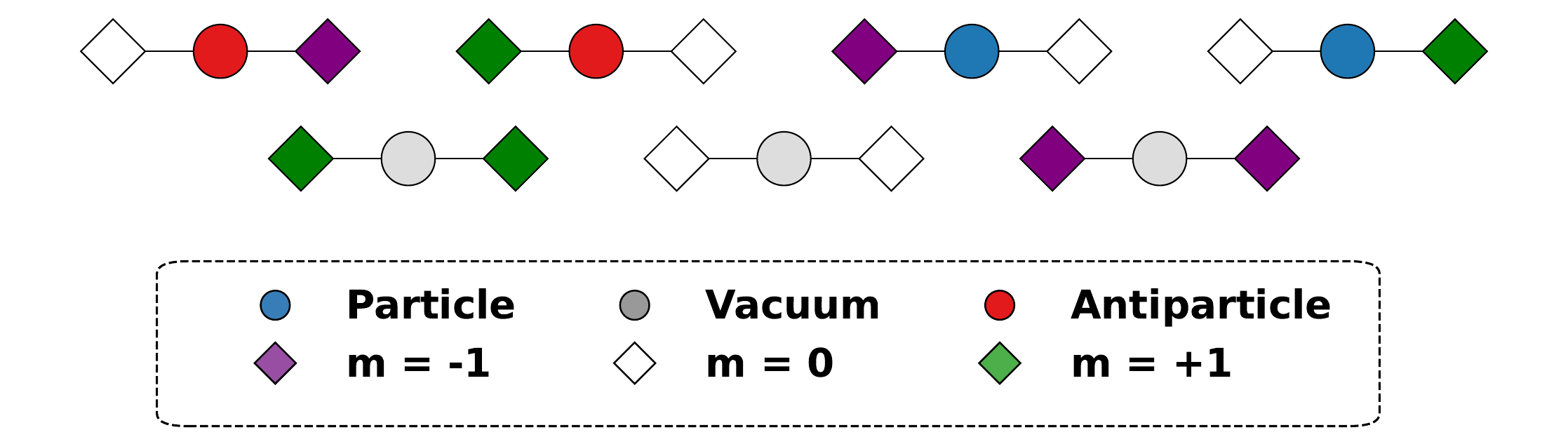}
    }
\subfloat[\scriptsize \sffamily\bfseries\textls{Qudit encoding}]{%
       \includegraphics[width=0.49\textwidth]{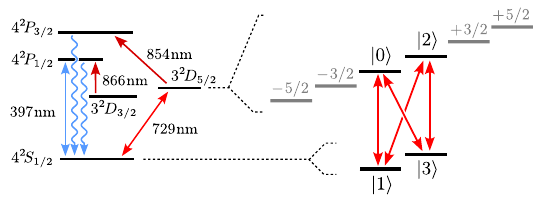}
    }
    \caption{Illustration of scattering dynamics in the lattice gauge theory: the matter sites are represented by circles and the gauge sites are represented by diamonds; panel~(a) depicts meson-antimeson collision, and panel~(b) depicts configurations allowed by Gauss's law (without a particle hole transformation $\hat{G}_j = \hat{s}^z_{j,j+1}-\hat{s}^z_{j-1,j}-\hat{\rho}_j$). (c) A simplified level diagram of a $^{40}\text{Ca}^+$ ion, a prominent choice of atom in state-of-the-art trapped-ion quantum processors. Coupling of the $4^2S_{1/2}$ ground states to the short-lived $4^2P$ states enables efficient laser cooling and state detection (blue, $397~\text{nm}$). Spontaneous decay to a metastable $3^2D_{3/2}$ state is suppressed by continuous re-pumping (red, $866~\text{nm}$). Due to the anomalous Zeeman effect a magnetic bias field splits the $4^2S_{1/2}$ ground state and the long-lived $3^2D_{5/2}$ into two, respectively six sub-states, which are well suited to encode a qudit; here we show a possible encoding of a ququart ($d=4$). The qudit is manipulated using coherent narrow-band laser pulses, enabling a pair-wise coupling of the states as indicated by the arrows.}
    \label{fig:lattice_scheme}
\end{figure*}

In a scattering event, two elementary or composite particles approach each other, interact through a collision, and evolve into a post-collision state. The quantum wave function of the full system encodes the entire process --- from the incoming state to the final outgoing state. While particle colliders provide detailed information about the asymptotic (late-time) states, accessing the real-time quantum dynamics during the interaction remains an outstanding challenge. Currently, classical methods relying on perturbative expansions or effective theories are used to reconstruct this dynamics from late-time observables \cite{QCD_review,Berges_review}. However, a fully \textit{ab initio} time-resolved description of scattering remains out of reach. Quantum simulators offer a promising route to filling this gap by enabling the direct study of out-of-equilibrium quantum dynamics in gauge theories \cite{Dalmonte_review, Zohar_review, Aidelsburger2011, Zohar_NewReview, klco2021standard, Bauer_ShortReview, Bauer_review, dimeglio2023quantum, Cheng_review, Halimeh_review, Cohen:2021imf, Lee:2024jnt, Turro:2024pxu,bauer2025efficientusequantumcomputers}. They offer time-resolved snapshots of the quantum dynamics, potentially revealing the microscopic details that govern the interaction cross-section. While the far-from-equilibrium nonperturbative nature of these dynamics leads to an exponential scaling of computational cost on classical devices, quantum simulators can, in principle, address such problems with polynomial resources by leveraging quantum advantage.

The fundamental framework for such HEP studies is given by gauge theories, which describe how elementary particles interact with force fields through gauge bosons~\cite{Gattringer_book}.
In practice, the lattice formulation of these theories, so-called lattice gauge theories (LGTs) have been used successfully to shed light on quark confinement \cite{Wilson1974} and other properties of quantum chromodynamics (QCD) \cite{Rothe_book}. However, while their investigation through Monte Carlo methods generated a lot of success \cite{Creutz1980MC,Montvay_book}, they are mostly limited to equilibrium studies due to the sign problem \cite{deforcrand2010simulatingqcdfinitedensity,Troyer2005computational}. Going beyond Monte Carlo, the use of tensor networks has enabled the study of LGT time-evolution dynamics, albeit limited to $1+1$D or very small systems in $2+1$D, and usually restricted to small evolution times due to the exponential growth of the bond dimension \cite{Uli_review,Orus2013,Paeckel_review,Orus2019}. In the last decade, quantum simulation experiments have emerged as a promising approach for investigating various HEP phenomena \cite{Martinez2016,Klco2018,Goerg2019,Schweizer2019,Mil2020,Yang2020,Wang2021,Su2022,Zhou2022,Wang2023,Zhang2023,Ciavarella2024quantum,Ciavarella:2024lsp,Farrell:2023fgd,Farrell:2024fit,zhu2024probingfalsevacuumdecay,Ciavarella:2021nmj,Ciavarella:2023mfc,Ciavarella:2021lel,Gustafson:2023kvd,Gustafson:2024kym,Lamm:2024jnl,Farrell:2022wyt,Farrell:2022vyh,Li:2024lrl,Zemlevskiy:2024vxt,Lewis:2019wfx,Atas:2021ext,ARahman:2022tkr,Atas:2022dqm,Mendicelli:2022ntz,Kavaki:2024ijd,Than:2024zaj,Angelides2025first,cochran2024visualizingdynamicschargesstrings,gyawali2024observationdisorderfreelocalizationefficient,gonzalezcuadra2024observationstringbreaking2,crippa2024analysisconfinementstring2}, motivating the development of more sophisticated quantum hardware and algorithms.

More recently, there has been great interest in quantum simulation of scattering processes. This has included theoretical proposals \cite{Surace2021scattering,su2024particlecollider,chai2025quantumsimulationmesonscattering,Bennewitz2025simulatingmeson}, quantum simulation experiments of scattering in non-LGT Hamiltonians \cite{Chai2025fermionicwavepacket,Zemlevskiy:2024vxt,farrell2025digitalquantumsimulationsscattering,ingoldby2025realtimescatteringquantumcomputers}, of wave-packet preparation in LGTs \cite{Farrell:2024fit,Davoudi2024scatteringwave}, and of scattering dynamics in simple LGTs with a two-level discretization of the gauge field \cite{schuhmacher2025observationhadronscatteringlattice,davoudi2025quantumcomputationhadronscattering}. Although far from the fully fledged gauge theories of the Standard Model, such toy model LGTs are incredibly important when it comes to the development of HEP quantum simulators. Furthermore, they have become popular venues for exploring exotic quantum many-body dynamics and properties relevant to condensed matter and quantum information \cite{Surace2020,Desaules2022weak,Desaules2022prominent,aramthottil2022scar,Smith2017,Brenes2018,smith2017absence,karpov2021disorder,Sous2021,Chakraborty2022,Halimeh2021enhancing,Tarabunga2023many,hartse2024stabilizerscars,Smith2025nonstabilizerness,Falcao2025Nonstabilizerness,Esposito2025magic,Desaules2024ergodicitybreaking,desaules2024massassistedlocaldeconfinementconfined,jeyaretnam2025hilbertspacefragmentationorigin,ciavarella2025generichilbertspacefragmentation}. Nevertheless, in order to truly make HEP quantum simulators \textit{bona fide} complementary venues to particle colliders when it comes to scattering experiments, an effort must be taken to (i) encode more levels of the gauge field in order to approach the Kogut--Susskind limit \cite{Buyens2014,Zache2021achieving}, (ii) implement non-Abelian gauge groups such as $\mathrm{SU}(3)$, and (iii) embed the LGT in $3+1$D.

Several proposals have recently emerged on the quantum simulation of LGTs using qudits \cite{ciavarella2022conceptualaspectsoperatordesign,Popov2024variational,Calajo2024digital,kürkçüoglu2024quditgatedecompositiondependence,ballini2025symmetryverificationnoisyquantum,gaz2025quantumsimulationnonabelianlattice,jiang2025nonabeliandynamicscubeimproving}. Encoding information in $d\geq2$ levels, qudits promise to be more naturally suited to LGT quantum simulation than their binary qubit counterparts~\cite{Ringbauer_2022, meth2025}. Indeed, gauge fields in nature have an infinite-dimensional Hilbert space, and encoding them into qubits leads to significant overhead in gate and circuit complexity. Encoding a high-dimensional gauge field into a high-dimensional qudit, on the other hand, provides for a natural and efficient implementation. This is particularly relevant for quantum simulations of scattering processes, where a multilevel representation of the gauge field is essential for increasing the number of possible post-collision particles, as we will further discuss in the following.

In this work, we consider a quantum link model (QLM) formulation \cite{Chandrasekharan1997,Wiese_review} of the lattice Schwinger model \cite{Coleman1975,Coleman1976}, where electric and gauge fields are represented by spin-$S$ $\hat{s}^z$ and spin ladder operators, respectively, and propose efficient qudit circuits to study scattering processes on them; see Fig.~\ref{fig:lattice_scheme}. To achieve optimal circuit depth, we employ qudits \cite{Ringbauer_2022} to realize the spin-$S$ operators. In particular, to realize the spin-$1$ QLM Hamiltonian efficiently, we encode the gauge degrees of freedom using qutrits, i.e., three-level qudits corresponding to electric field eigenstates $m = \{-1, 0, +1\}$. This native qudit encoding allows for the direct implementation of the spin operators $\hat{s}^z$ and $\hat{s}^+$ without the overhead of decomposing into multiple two-level systems. To implement scattering dynamics, we prepare two particle-antiparticle pairs localized at opposite ends of a sub-chain bounded by two confining walls. These walls prevent motion away from the central region, forcing the pairs to accelerate towards one another. The walls are held fixed during an initial holding period, chosen to match the approximate time it takes for the pairs to reach the center of the chain. We observe that scattering occurs only when the gauge-coupling strength~$g$ is sufficiently large; otherwise, the particles effectively pass through each other without interaction. To confirm that the observed dynamics arise from genuine interactions, we compare the full evolution to that of single pairs propagating in isolation and subtract the resulting charge profile.

The remainder of the paper is structured as follows. Section \ref{ModelWithMatter} provides a brief overview of $1+1$D lattice QED, following which we describe the procedure of integrating out matter in Sec.~\ref{Sec:IntOut}. Section \ref{SecLinkCircuit} details the qudit-based quantum circuit for the resulting matter-integrated-out model, detailing the Trotterization approach, gate decomposition, and noise modeling. Section \ref{SecDynamics} explores meson-meson and meson-antimeson scattering dynamics, comparing noisy simulations with exact results. In Sec.~\ref{circ_with_mat}, we construct the circuits for the original model, retaining matter fields. Section \ref{NoiseComp} investigates noise resilience for both the circuits, demonstrating improved performance in the matter-integrated-out approach due to reduced gate counts and Hilbert space leakage. Section~\ref{Sec:experimental} outlines the experimental trapped-ion platform for implementation. We conclude and provide outlook in Sec.~\ref{SecConOut}.

\section{The Model}
\label{ModelWithMatter}
In this work, we adopt a QLM formulation \cite{Chandrasekharan1997,Wiese_review} of $1+1$D QED. To make the model amenable to quantum simulation, we map the fermionic matter fields to hardcore bosons via the Jordan--Wigner transformation \cite{altland2010condensed}, allowing the matter fields to be naturally encoded in qubits. The resulting Hamiltonian is given by \cite{osborne2023spinsmathrmu1quantumlink,osborne2023probingconfinementdynamicalquantum,Kasper2017Feb,Hauke2013Nov}

\begin{align}
	\hat{H} =&     \hat{H}^{\text{min}}  + \hat{H}^{\text{sg}}\notag\\
        =&\sum_{j = 0}^{L-2} \kappa (\hat{\sigma}_j^+ \hat{s}^+_{j, j+1} \hat{\sigma}^-_{j+1} + \text{H.c.})\notag\\
        &+ \sum_{j = 0}^{L-1} \mu (-1)^j \hat{\sigma}_j^z + \frac{g^2}{2} \sum_{j = 0}^{L-2} \big(\hat{s}^z_{j,j+1}\big)^2,
\end{align} 
where $\hat{\sigma}_j^{\pm}$ are raising and lowering operators at site $j$ representing matter annihilation and creation operators, and $\hat{s}^{\pm}_{j,j+1}, \hat{s}^z_{j,j+1}$ are spin-$S$ operators representing the gauge and electric fields on the links, respectively. 

The Hamiltonian consists of two parts: the minimal coupling term $\hat{H}^\text{min}$, with a coupling strength $\kappa$, and $\hat{H}^{\text{sg}}$, which contains both the mass and the gauge coupling terms. Here, $g$ is the strength of the gauge coupling, and the staggered mass term $\mu(-1)^j$, involving the operator $\hat{\sigma}^z_j$ that represents the matter occupation at site $j$, differentiates between matter and antimatter on the lattice, with odd sites representing matter (particles carrying negative charge) and even sites representing antimatter (particles carrying positive charge).

The generator of the $\mathrm{U}(1)$ gauge symmetry is expressed as $\hat{G}_j = \hat{s}^z_{j,j+1}-\hat{s}^z_{j-1,j}-\hat{\rho}_j$, where $\hat{\rho}_j$ is the charge operator defined as
    \begin{equation}
	\hat{\rho}_{j} = \frac{\hat{\sigma}^z_{j}+(-1)^j}{2}. \label{chargej}
\end{equation}
The physical sector is spanned by states $\{\ket{\Psi}\}$ satisfying $\hat{G}_j \ket{\Psi} = 0,\,\forall j$: this condition can be viewed as a discretized version of Gauss's law, imposing an intrinsic relation between the matter occupation on site $j$ and the allowed electric-field configurations on its neighboring links, as shown in Fig.~\ref{fig:lattice_scheme}(b).

To eliminate the staggered mass and make pair creation explicit, we perform a particle-hole (p-h) transformation at the spin level, acting only on operators at odd sites $j$ as \cite{Hauke2013Nov}
\begin{equation}
    \hat{\sigma}^-_{j} \to \hat{\sigma}^+_{j}, \, \hat{s}^-_{j,j+1} \to \hat{s}^+_{j,j+1}, \label{ph_mapping}
\end{equation}
while leaving the operators at even sites unchanged. This transformation leads to the Hamiltonian terms
\begin{subequations}
\begin{align}
    \hat{H}^{\text{min}} &= \sum_{j = 0}^{L-2}\kappa  \big(\hat{\sigma}^+_{j}\hat{s}^+_{j,j+1}\hat{\sigma}^+_{j+1}+\text{H.c.}\big) \label{minCouple2}\\
    \hat{H}^{\text{sg}} &= \frac{\mu}{2}\sum_{j=0}^{L-1} \hat{\sigma}^z_j+\frac{g^2}{2}\sum_{j=0}^{L-2}\big(\hat{s}^z_{j,j+1}\big)^2\label{non-int}
\end{align}
\end{subequations}
along with the particle-hole transformed gauge generators 
\begin{equation}
	\hat{G}_j^{\text{p-h}} = (-1)^j\left[\hat{s}^z_{j-1,j} + \hat{s}^z_{j,j+1}-\frac{\hat{\sigma}^z_{j}+1}{2}\right]. \label{gausslawIntOut}
\end{equation}

In our analysis, we restrict to the spin-$1$ representation, where each link is encoded in a qudit. The intrinsic multilevel structure of qudits \cite{Mato2024Oct,Ringbauer_2022} naturally lends itself to encoding links as qutrits, offering a more efficient representation than using two qubits. The operators on the link ($j,j+1$) are then represented as
\begin{equation}
    \hat{s}^{z}_{j,j+1} =  \ketbra{0}-\ketbra{2}, \quad \hat{s}^{+}_{j,j+1} = \sqrt{2} \big[\ketbra{2}{1}+\ketbra{1}{0}\big],
\end{equation}
where the basis states $\{\ket{0}, \ket{1}, \ket{2}\}$ span the local qutrit Hilbert space. These states are defined as
\begin{align}
    \ket{0} &= \ket{m = 1} \notag\\
    \ket{1} &= \ket{m = 0} \notag\\
    \ket{2} &= \ket{m = -1}, \label{quantum_reg_state}
\end{align}
where $m$ is the eigenvalue of $\hat{s}^z$. This qutrit encoding is realized in three levels of ${}^{40}\text{Ca}^+$ ions, as depicted in Fig.~\ref{fig:lattice_scheme}(c).

\subsection{Matter-integrated-out formulation\label{Sec:IntOut}}
The gauge symmetry of the QLM enables further simplification of the model, which we now discuss in detail. Using the minimal-coupling term in the particle-hole transformed QLM~\eqref{minCouple2}, we identify only two allowed processes for a configuration of two matter sites connected by a single link within the physical sector. We define the qubit states encoding matter occupation as $\ket{\uparrow}$ for occupied sites and $\ket{\downarrow}$ for unoccupied sites, while the link encoding follows the definition in Eq. \eqref{quantum_reg_state}. The first process involves the vacuum state $\ket{1\downarrow 1 \downarrow 1} \longleftrightarrow \ket{1\uparrow 0 \uparrow 1}$, captured by the operator $\hat{P}_{j-1,j}^{1} \hat{\sigma}^{x;01}_{j,j+1} \hat{P}_{j+1,j+2}^{1}$, where $\hat{P}_{j-1,j}^{1}$ projecting the qudit encoding the link to the left onto the state $\ket{1}$. Here, $\hat{\sigma}^{\mu; ij}$ denotes the generalized Pauli operators acting on the two-dimensional subspace $\{\ket{i}, \ket{j}\}$ where $\mu \in 0, x, y, z$. 
In terms of only the electric-field degrees of freedom, this first process connects 

\begin{align}
(m_{j-1,j},m_{j,j+1},m_{j+1,j}): (0,0,0) \longleftrightarrow (0,1,0).
\end{align}
The second process involves the transition $\ket{0\uparrow 1 \uparrow 0} \longleftrightarrow \ket{0\downarrow 2 \downarrow 0}$, which corresponds to
\begin{align}
(m_{j-1,j},m_{j,j+1},m_{j+1,j+2}): (1,0,1) \longleftrightarrow (1,-1,1),
\end{align}
and is captured by $\hat{P}_{j-1,j}^{0} \hat{\sigma}^{x;12}_{j,j+1} \hat{P}_{j+1,j+2}^{0}$.

Without loss of generality, we set the links at the edges of the chain to be in the $m = 0$ state. Thus, the minimal coupling terms are
\begin{subequations}
\begin{align}
    \hat{H}^{\text{min}}_{\text{links};j=0} &= \sqrt{2}\kappa \hat{\sigma}^{x;01}_{0,1} \hat{P}_{1,2}^{1},\\
    \hat{H}^{\text{min}}_{\text{links};j=L-2} &= \sqrt{2} \kappa \hat{P}_{L-3,L-2}^{0} \hat{\sigma}^{x;01}_{L-2,L-1},
\end{align}
on the edges, while the bulk terms ($j\neq 0,L-2$) are given by
\begin{align}
        \hat{H}^{\text{min}}_{\text{links};j} =& \sqrt{2}\kappa \hat{P}_{j-1,j}^{1} \hat{\sigma}^{x;01}_{j,j+1} \hat{P}_{j+1,j+2}^{1}\notag\\ &+ \sqrt{2}\kappa\hat{P}_{j-1,j}^{0} \hat{\sigma}^{x;12}_{j,j+1} \hat{P}_{j+1,j+2}^{0}.
\end{align}
\end{subequations}
The full matter-integrated-out Hamiltonian is then
\begin{align}
		\hat{H}_{\text{links}} &= \hat{H}^{\text{min}}_{\text{links}} + \hat{H}^{\text{sg}}_{\text{links}}\notag\\
        &=\sum_{j=0}^{L-2} \hat{H}^{\text{min}}_{\text{links};j}+ 2\mu\sum_{j=1}^{L-2} \hat{s}^z_{j,j+1}+\frac{g^2}{2}\sum_{j=0}^{L-2}\big(\hat{s}^z_{j,j+1}\big)^2,
        \label{h_qudit_qlm}
\end{align}
where the middle term is obtained using Gauss's law $\hat{G}^\text{p-h}_j=0$ to obtain $\sum_j\hat{\sigma}^z_j=4\sum_j\hat{s}^z_{j,j+1}$ up to an inconsequential energy constant that we omit.
This resulting Hamiltonian is rather similar to that of Refs.~\cite{Desaules2022prominent} and \cite{Popov2024Feb}, but with some slight differences owing to our use of open rather than periodic boundary conditions.

\begin{figure*}
\captionsetup[subfloat]{position=top,justification=raggedright,singlelinecheck=false}
  \centering
  \subfloat[]{%
    \resizebox{\linewidth}{!}{%
    \begin{quantikz}[row sep={1.5cm,between origins}]     
\lstick{$\ket{\text{qudit}_i}$} & \gate[2][2cm][1cm]{\text{CRX}^{c, ab}_{ij}(T)} & \midstick[2,brackets=none]{\textbf{\large =}} & & & \ctrl{1} & & \ctrl{1} & & & \\     
\lstick{$\ket{\text{qudit}_j}$} & & & \gate[1][1cm][1cm]{\text{H}^{a,b}} & \gate[1][1cm][1cm]{\text{RZ}^{a,b}\big(\frac{T}{2}\big)} & \gate[1][1cm][1cm]{\text{CX}_{c, a\leftrightarrow b}} & \gate[1][1cm][1cm]{\text{RZ}^{a,b}\big(-\frac{T}{2}\big)} & \gate[1][1cm][1cm]{\text{CX}_{c, a\leftrightarrow b}} & \gate[1][1cm][1cm]{\text{H}^{a,b}} & & \\     
\end{quantikz}
}%
}
  \hfill
  \subfloat[]{%
    \resizebox{\linewidth}{!}{%
    \begin{quantikz}[row sep={2.7cm,between origins}, font=\Huge]
    \lstick{$\ket{\text{qutrit}_{j-1,j}}$} & \gate[3][3cm][1cm]{\hat{U}^j_{min} (T)} & & \midstick[3,brackets=none]{\textbf{\Huge =}} & & \ctrl{2} & &\ctrl{2} & \ctrl{2} & &\ctrl{2} & \\
    \lstick{$\ket{\text{ququart}_{j,j+1}}$} & & & & & & \gate[2][2cm][2cm]{\text{CRX$^{3, 12}_{(j,j+1), (j+1,j+2)}$}(T)} & & & \gate[2][2cm][2cm]{\text{CRX$^{3,01}_{(j,j+1), (j+1,j+2)}$}(T)} & &\\
    \lstick{$\ket{\text{ququart}_{j+1,j+2}}$} & & & & & \gate[1][2cm][2cm]{\text{CX}_{0, 0\leftrightarrow 3}} &   & \gate[1][2cm][2cm]{\text{CX}_{0, 0\leftrightarrow 3}} & \gate[1][2cm][2cm]{\text{CX}_{1, 1\leftrightarrow 3}} &   & \gate[1][2cm][2cm]{\text{CX}_{1, 1\leftrightarrow 3}} &\\
\end{quantikz}
}
}
\caption{(a) The subcircuit $\text{CRX}^{c, ab}_{ij}(T)$ applies a controlled-$\hat{R}_x^{ab}(T)$ operation on qudit $j$ by an angle $T$ within the appropriate subspaces, depending on which projector condition is satisfied. $\mathrm{H}^{i,j}$ gate acts as the Hadamard gate in the $\{\ket{i}, \ket{j}\}$ subspace. (b) The full circuit for $\hat{U}_j^{\text{min}}$ - Different conditional $\hat{R}_x^{ab} (T)$ rotations are applied on the qudit at link $(j, j+1)$ conditioned on whichever projector condition is satisfied on both the neighboring qudits.}\label{fig:min_circuit}
\end{figure*}
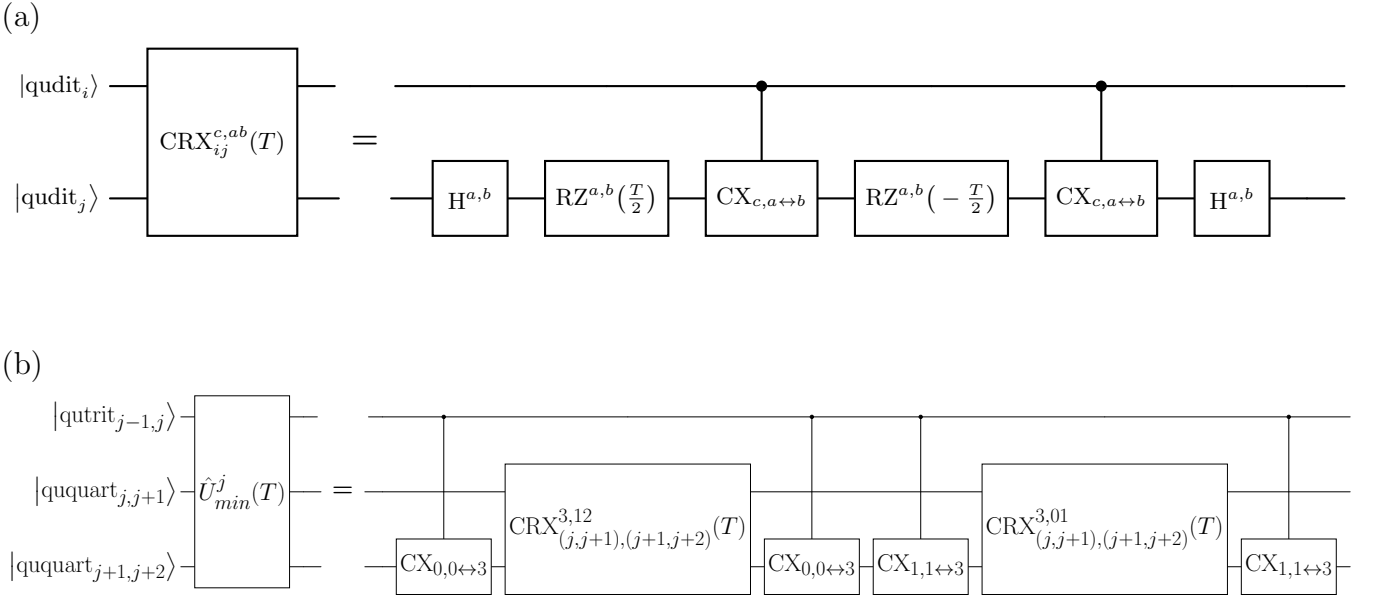

\section{The matter-integrated-out circuit}
\label{SecLinkCircuit}
In the following, we demonstrate explicit circuit constructions to simulate the dynamics under the Hamiltonian \eqref{h_qudit_qlm}. For this, we employ a second-order Trotter decomposition
\begin{equation}
	\hat{U}_{\mathrm{ST}}(t) = \Big(\prod_j \underbrace{e^{-\imath \frac{T}{2} \hat{H}^{\text{sg}}_{\text{links};j}}}_{\displaystyle \hat{U}^{\text{sg}}_{\text{links};j} (T/2)} \, \underbrace{e^{-\imath T  \hat{H}^{\text{min}}_{\text{links};j }}}_{\displaystyle \hat{U}^{\text{min}}_{\text{links};j} (T)} e^{-\imath \frac{T}{2}  \hat{H}^{\text{sg}}_{\text{links};j}}\Big)^N,
    \label{2ndorderTrotterization_mio}
\end{equation}
where the Trotter step size is defined as $T \equiv t/N$.
We present the circuits for implementing each term in $\hat{U}_{\mathrm{ST}}(t)$ below. The first operator $\hat{U}^{\text{sg}}_{\text{links};j} (T)$ can be easily implemented using virtual $RZ$ gates. For qudits, these gates generalize to
\begin{equation}
\text{VRZ}_{j}^{a}(\phi) = e^{-\imath\phi \ketbra{a}_{j,j+1}}. \label{virz}
\end{equation}
These ``virtual'' gates are realized through classical frame adjustments rather than physical pulses and are therefore noiseless.

\begin{figure*}[t]
\captionsetup[subfloat]{position=top,justification=raggedright,singlelinecheck=false, font=large}
  \centering
  \subfloat[]{%
    \resizebox{1.0\textwidth}{!}{%
        \centering
    	\includegraphics[width=\linewidth]{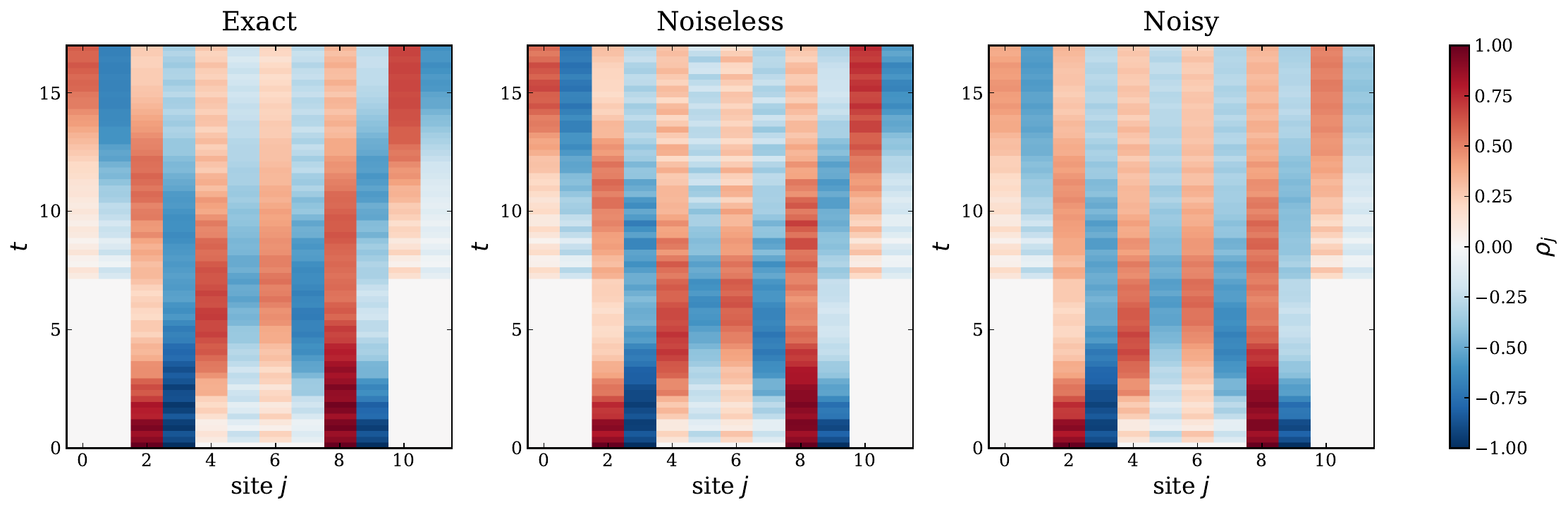}
        }
        }\hfill
         \subfloat[]{\resizebox{0.38\textwidth}{!}{%
        \centering  	\includegraphics[width=\linewidth]{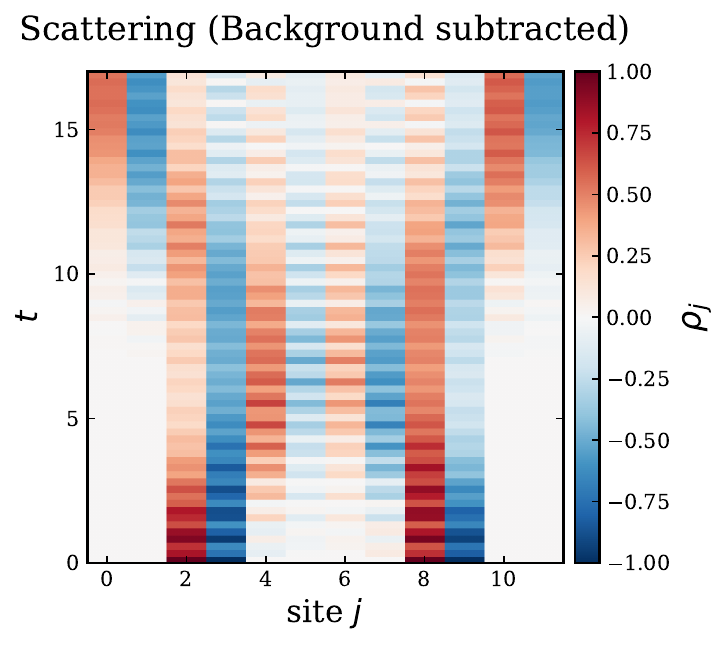}
        }
        }
        \subfloat[]{%
    \resizebox{0.38\textwidth}{!}{%
        \centering
    	\includegraphics[width=\linewidth]{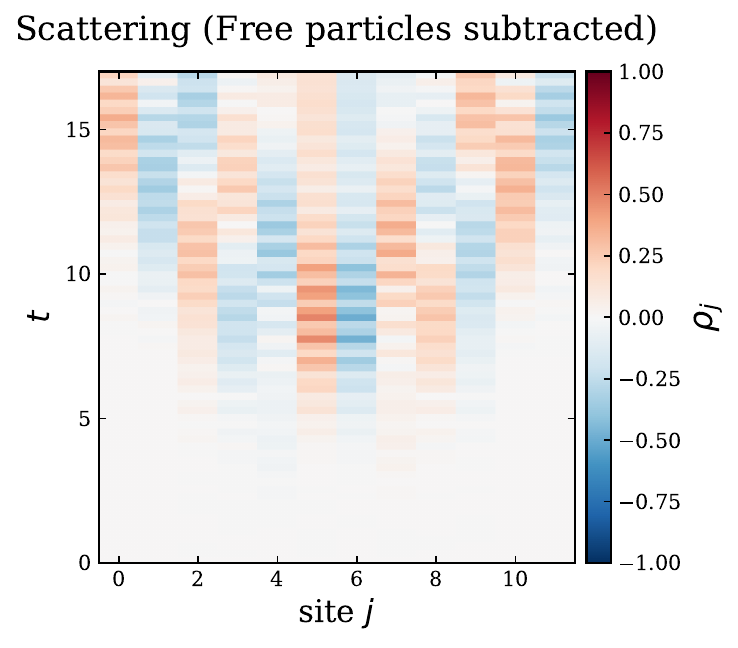}
        }
        }
        \caption{Charge density evolution for meson-meson collision with $\mu=1$, $g=3$, $\kappa=1$, Trotter step size $T= 0.25$. The error probabilities are $p_{\text{one-body}} = 3\times 10^{-6}$ and $p_{\text{CX}} = 2 \times 10^{-4}$. In color is the expectation value of the local charge operators defined in \eqref{chargej}: (a) Comparison of the exact, noiseless, and noisy circuit simulations. (b) Charge dynamics after vacuum fluctuations are subtracted off, and (c) Scattering signature obtained by subtracting off the free particles. \label{meson-meson_g=3}}
\end{figure*}

The circuit implementing $\hat{U}^{\text{min}}_{\text{links};j}$ is shown in Fig.~\ref{fig:min_circuit}. To demonstrate the validity of this construction, we perform an analytical decomposition of $\hat{U}^{\text{min}}_{\text{links};j}$, where we use the identities
\begin{align}
(\hat{P}^{0}_{j-1, j} \hat{\sigma}^{x;12}_{j, j+1} \hat{P}^{0}_{j+1, j+2})^2 & = \hat{{I}}_{\hat{P}^0_{j,j+1}} \hat{{I}}^{1,2}_{j, j+1} \hat{{I}}_{\hat{P}^0_{j+1, j+2}} =  \hat{{I}}',\nonumber\\
(\hat{P}^{1}_{j-1, j} \hat{\sigma}^{x;01}_{j, j+1}  \hat{P}^{1}_{j+1, j+2})^2 & = \hat{{I}}_{\hat{P}^{1}_{j-1, j}} \hat{{I}}^{0,1}_{j, j+1} \hat{{I}}_{\hat{P}^{1}_{j+1, j+2}} = \hat{{I}}''.
\end{align}
Here, each $\hat{I}$ represents the identity operator restricted to the relevant subspace defined by either the projectors or $\hat{\sigma}^{x;ij}$ operator acting on the relevant qudits. Thus, $\hat{I}'$ and $\hat{I}''$ act as identity operators over the respective joint subspaces. Consequently, combining these results, we have $(\hat{H}_{\text{links};j}^{\text{min}})^2 = \hat{{I}}' + \hat{{I}}'' \equiv \hat{{{I}}}$ where $\hat{I}$ denotes the identity operator on the full combined subspace.
This can be used to facilitate the decomposition
\begin{align}
e^{-\imath T \hat{H}_{\text{links};j}^{\text{min}}} & =  \sum_{n=0}^\infty\frac{\big({-}\imath T \hat{H}_{\text{links};j}^{\text{min}}\big)^n}{n!} \nonumber\\
& = \mathbb{1} - \hat{{{I}}} + \cos(T) \,\hat{{{I}}} - \imath \sin(T) \,\hat{H}_{\text{links};j}^{\text{min}} \nonumber\\
& = \mathbb{1} - \hat{{{I}}} + \hat{P}^1_{j-1,j}\, e^{-\imath \sqrt{2}\kappa T \hat{\sigma}^{x;01}_{j, j+1}}\, \hat{P}^1_{j+1,j+2} \nonumber\\
& \quad{} + \hat{P}^{0}_{j-1,j}\, e^{-\imath \sqrt{2} \kappa T \hat{\sigma}^{x;12}_{j, j+1}}\, \hat{P}^{0}_{j+1,j+2}.
\end{align}
Through this decomposition, it becomes apparent that the circuit should leave those three-body states unaffected that do not survive the action of $\hat{H}_{\text{links};j}^{\text{min}}$. For the states affected by the circuit, their conditional dynamics is decided by one of the PXP-type terms of the form $\hat{P}^m_{j-1,j}\, e^{-\imath \sqrt{2} \kappa T \hat{\sigma}^{x;ab}_{j, j+1}}\,\hat{P}^m_{j+1,j+2}$. Each of these terms corresponds to a conditional $\hat{R}_x^{ab}(T) \equiv \text{exp}({-\imath \kappa T\hat{\sigma}^{x;ab}})$ operation applied on the qudit at the link $(j, j+1)$ based on the projector conditions satisfied by the neighboring qudits at links $(j-1,j)$ and $(j+1, j+2)$.

The circuit construction proceeds as follows. For the first PXP-type term, the joint state $\ket{1_{j-1,j}1_{j+1,j+2}}$ satisfies the projector conditions and is mapped to $\ket{1_{j-1,j}3_{j+1,j+2}}$. A controlled rotation operation $\hat{R}_x^{01}(T)$ is then applied in the subspace $\{\ket{0}, \ket{1}\}$ with the qudit at link $(j+1,j+2)$ serving as the control in state $\ket{3}$. The mapping is subsequently reversed. Similarly, for the second PXP-type term, the joint state $\ket{0_{j-1,j}0_{j+1,j+2}}$ is mapped to $\ket{0_{j-1,j}3_{j+1,j+2}}$, followed by a controlled rotation $\hat{R}_x^{12}(T)$ in the subspace $\{\ket{1}, \ket{2}\}$ with the qudit at link $(j+1,j+2)$ serving as the control in state $\ket{3}$, and then the mapping is reversed. This construction ensures that the circuit can be implemented using only $8$ entangling gates (Fig.~\ref{fig:min_circuit}(a)), for which we use controlled exchange (CX) gates \cite{Ringbauer_2022}, denoted as $\mathrm{CX}_{c, l_1 \leftrightarrow l_2}$ which apply an $X$ gate between the $\ket{l_1}$ and $\ket{l_2}$ states of the target qudit, conditioned on the control qudit being in state $\ket{c}$,
\begin{equation}\label{CX}
\begin{aligned}
\operatorname{CX}_{c, l_1 \leftrightarrow l_2}:\left\{
\begin{array}{l}
\left|c, l_1\right\rangle \leftrightarrow \left|c, l_2\right\rangle \\
|j, k\rangle \rightarrow |j, k\rangle \quad \text{for } j \neq c, \, k \neq l_1, l_2.
\end{array}
\right.
\end{aligned}
\end{equation}

\subsection{Noise estimation}
In our simulations, we consider the trapped-ion architecture described in \cite{Ringbauer_2022} as the experimental platform, which will be discussed briefly in Sec.~\ref{Sec:experimental}. To model the dynamics theoretically, we use the Munich Quantum Toolkit \cite{mqt}, specifically the \texttt{mqt.qudits} package, which is tailored to simulate qudit-based systems using the native gate set available on trapped-ion hardware. 
    
For this platform, the estimated error rates are $p_{\text{one-body}} = 3\times10^{-5-\alpha}$ for single-qudit (one-body) gates and $p_{\text{two-body}} \approx 10^{-3-\alpha}$ for two-qudit (two-body) gates, assuming that depolarization and dephasing occur with equal probability. Specifically, the CX gates have error rates $p_{\text{CX}} = 2 \times 10^{-3-\alpha}$ and the Mølmer-Sørensen (MS)\cite{S_rensen_2000} gates have error rates $p_{\text{MS}} = 5 \times 10^{-3-\alpha}$. The parameter $\alpha \in [0,1]$ captures the variation between current and near-term achievable noise rates. In our analysis, we set $\alpha=1$.

\section{Scattering dynamics}
\label{SecDynamics}
In this section, we investigate the scattering dynamics using our proposed circuits on a qudit-based quantum processor. In particular, we consider meson-meson and meson-antimeson scattering processes. To assess the reliability and feasibility of our quantum simulations under realistic conditions, we compare the results obtained from noisy simulations to those obtained from noiseless, ideal simulations. 

To simulate these collisions, we begin by preparing the appropriate initial states, as illustrated in Fig.~\ref{fig:lattice_scheme}(a). We impart opposing momenta on the two composite particles by removing the minimal coupling terms for a finite holding period on the two sites immediately to the left (right) of the left (right) pair --- effectively placing them just beyond static potential walls --- following the scheme proposed in Ref.~\cite{su2024particlecollider} for a cold-atom quantum simulator, and recently realized on a digital IBM quantum computer \cite{schuhmacher2025observationhadronscatteringlattice} for a $1+1$D spin-$1/2$ $\mathrm{U}(1)$ QLM. In the matter-integrated-out picture, this corresponds to initializing the qudits inside the walls in the state $\ket{0}$ if they are adjacent to the walls, and in the state $\ket{1}$ otherwise. All qudits beyond the walls are initialized in the state $\ket{1}$.

\begin{figure*}[t]
\captionsetup[subfloat]{position=top,justification=raggedright,singlelinecheck=false, font=large}
  \centering
  \subfloat[]{%
    \resizebox{1.02\textwidth}{!}{%
        \centering
    	\includegraphics[width=\linewidth]{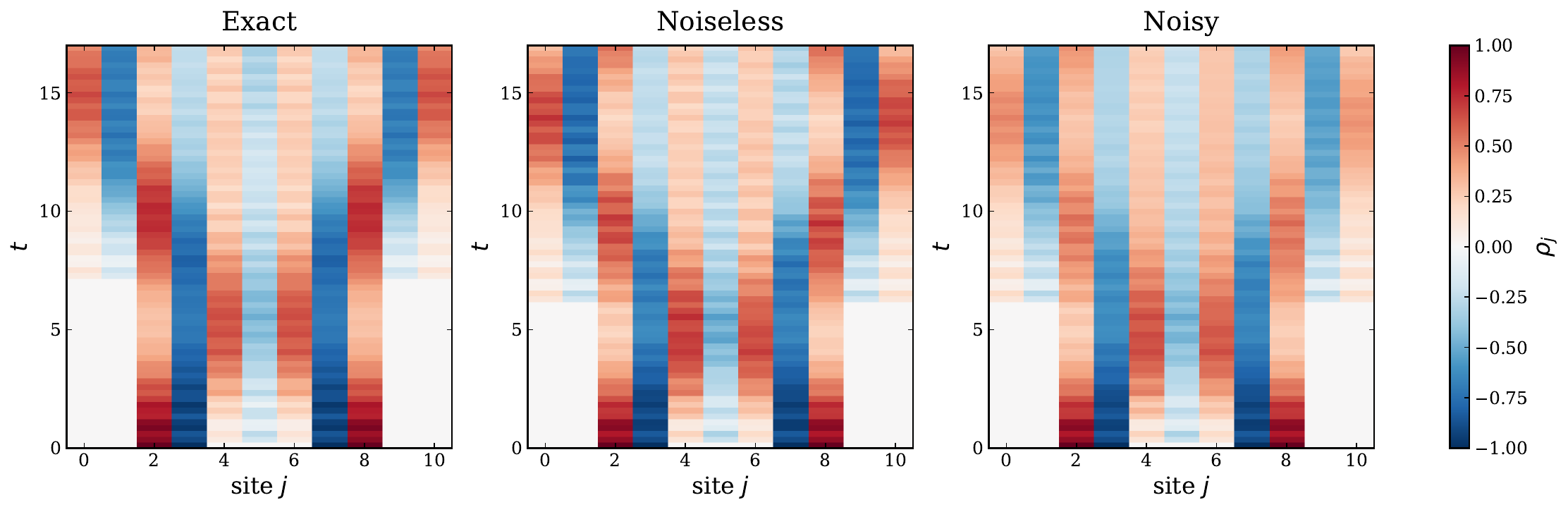}
        }
        }\hfill
        \subfloat[]{%
    \resizebox{1.02\textwidth}{!}{%
        \centering
    	\includegraphics[width=\linewidth]{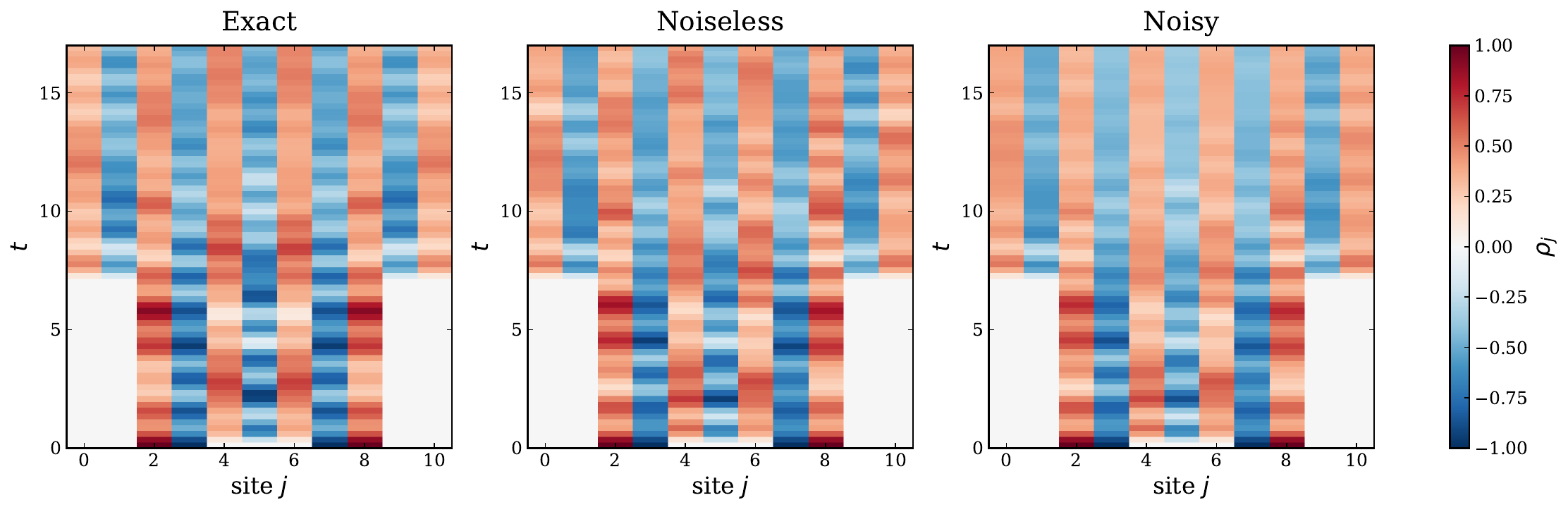}
        }
        }
        \caption{Charge density evolution for meson-antimeson collision comparing exact, noiseless, and noisy circuit simulations with $T = 0.25$ and parameters $\mu=1$, $\kappa=1$, (a) $g=3$, and (b) $g=0.5$. The error probabilities are $p_{\text{one-body}} = 3\times 10^{-6}$ and $p_{\text{CX}} = 2\times 10^{-4}$. In color is the expectation value of the local charge operators defined in \eqref{chargej}. \label{meson_antimeson}}
\end{figure*}

\begin{figure*}[t]
  \centering
    \captionsetup[subfloat]{position=top,justification=raggedright,singlelinecheck=false, font=large}
  \centering
  \subfloat[]{%
    \resizebox{0.37\textwidth}{!}{%
        \centering
    	\includegraphics[width=\linewidth]{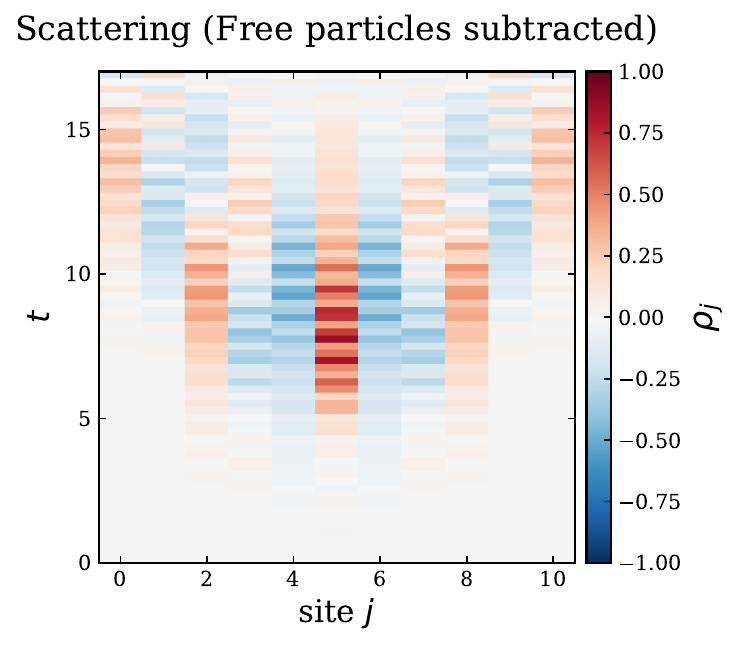}
        }
        }
         \subfloat[]{
         \resizebox{0.37\textwidth}{!}{%
        \centering  	\includegraphics[width=\linewidth]{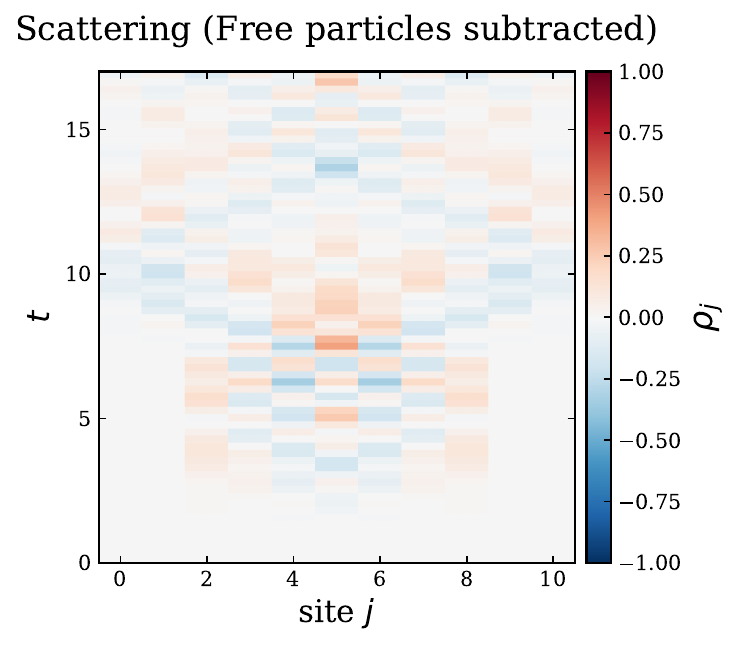}
        }
        }
        \caption{Scattering signature obtained by subtracting off the free particles with the model parameters $\mu=1$, $\kappa=1$, and Trotter step size $T= 0.25$: (a) $g=3$, and (b) $g=0.5$}
        \label{meson_antimeson_subtracted}
\end{figure*}

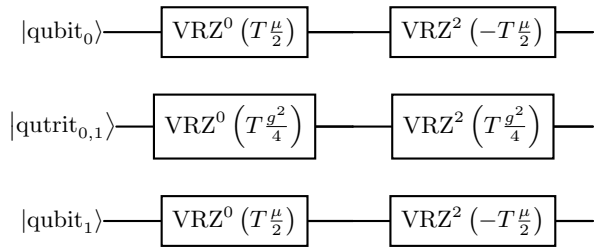
\begin{figure}[t]
    \begin{center}
    \begin{quantikz}
\ket{\text{qubit}_0}& \gate{\text{VRZ}^{0}\left(T\frac{\mu}{2}\right)} & \qw & \gate{\text{VRZ}^{2}\left(-T\frac{\mu}{2}\right)} &\qw\\
\ket{\text{qutrit}_{0,1}} & \gate{\text{VRZ}^{0}\left(T\frac{g^2}{4}\right)} &\qw & \gate{\text{VRZ}^{2}\left(T\frac{g^2}{4}\right)} &\qw \\
\ket{\text{qubit}_{1}} & \gate{\text{VRZ}^{0}\left(T\frac{\mu}{2}\right)} & \qw & \gate{\text{VRZ}^{2}\left(-T\frac{\mu}{2}\right)} &\qw
\end{quantikz}
\end{center}
    \caption{The noninteracting contribution to the quantum circuit with matter for a single link.}
    \label{fig:enter-label-2}
\end{figure}

\subsection{Meson-meson collision}
We begin by presenting the results for meson-meson collisions in a chain consisting of $12$ matter sites. In Fig.~\ref{meson-meson_g=3}(a), we compare the exact dynamics with those obtained from noiseless and noisy circuit simulations, where only the gauge degrees of freedom were simulated explicitly. The local charge at each matter site is then reconstructed from the electric fluxes using Gauss's law. 

As mentioned, the walls are kept turned on initially, confining the mesons within their boundaries, beyond which all matter sites remain in the vacuum state. Each meson is positioned next to one of the walls, with three vacuum matter sites separating them. The walls are then turned off at the moment of collision to allow the wave function to propagate beyond where the walls were.

We perform the simulations using a Trotter step size of $T = 0.25$, setting the model parameters $\kappa = 1$, $\mu = 1$, and $g=3$. The middle panel of Fig.~\ref{meson-meson_g=3}(a) depicts the results of the noiseless quantum simulation, which agree well with the exact time evolution. The right panel depicts the results when realistic, near-term quantum hardware noise is included. Although the noise gradually degrades the fidelity to the exact dynamics, the key features remain qualitatively visible, especially at early and intermediate times.

The observed dynamics reveal that the mesons travel towards the center of the chain, collide and bounce off each other, and subsequently spread at later times beyond the original location of the walls.
As the initial vacuum product state is not an eigenstate for finite $\mu$, we will observe some fluctuations in the background due to spontaneous pair creation and annihilation.
To better observe the scattering dynamics, we plot in Fig.~\ref{meson-meson_g=3}(b) the charge density from the scattering simulation with the background fluctuations subtracted off, found by simulating the same setup with no particles placed in the initial state.
Furthermore, in Fig.~\ref{meson-meson_g=3}(c), we plot the difference of the scattering simulation with two free meson simulations (one for the right-moving meson and one for the left-moving meson), all with background fluctuations subtracted off.
This allows us to clearly identify the occurrence of scattering: While the difference is zero before the collision time, the fact that it becomes nonzero afterwards is a clear signature of nontrivial interaction during the collision event.
Thus, for these chosen parameters, the mesons interact and reflect off each other rather than behaving as noninteracting, freely propagating excitations. 

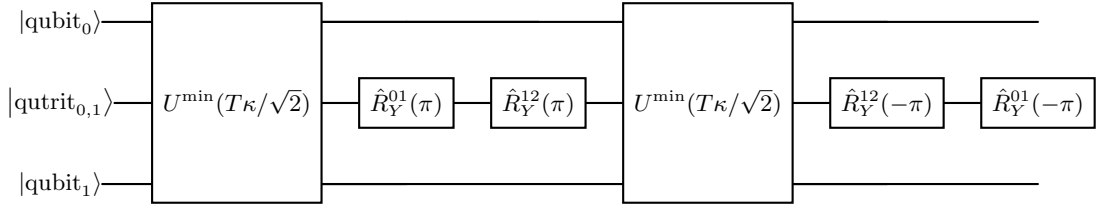
\begin{figure*}
\begin{center}
\centering
    \begin{quantikz}
\ket{\text{qubit}_0}&\gate[3,disable auto
height]{U^{\text{min}}(T\kappa/\sqrt{2})} &\qw &\qw &\gate[3,disable auto
height]{U^{\text{min}}(T\kappa/\sqrt{2})} & \qw &\\
\ket{\text{qutrit}_{0,1}} & & \gate{\hat{R}^{01}_{Y}(\pi)} & \gate{\hat{R}^{12}_{Y}(\pi)} & & \gate{\hat{R}^{12}_{Y}(-\pi)} &\gate{\hat{R}_{Y}^{01}(-\pi)}\\
\ket{\text{qubit}_{1}} & &\qw & & & \qw &
\end{quantikz}
\end{center}
\caption{The circuit for implementing a single minimal coupling term \eqref{MSdecomp}. \label{fig:enter-label2}}
    
\end{figure*}

\begin{figure*}
	\centering
	\includegraphics[width=1.02\linewidth]{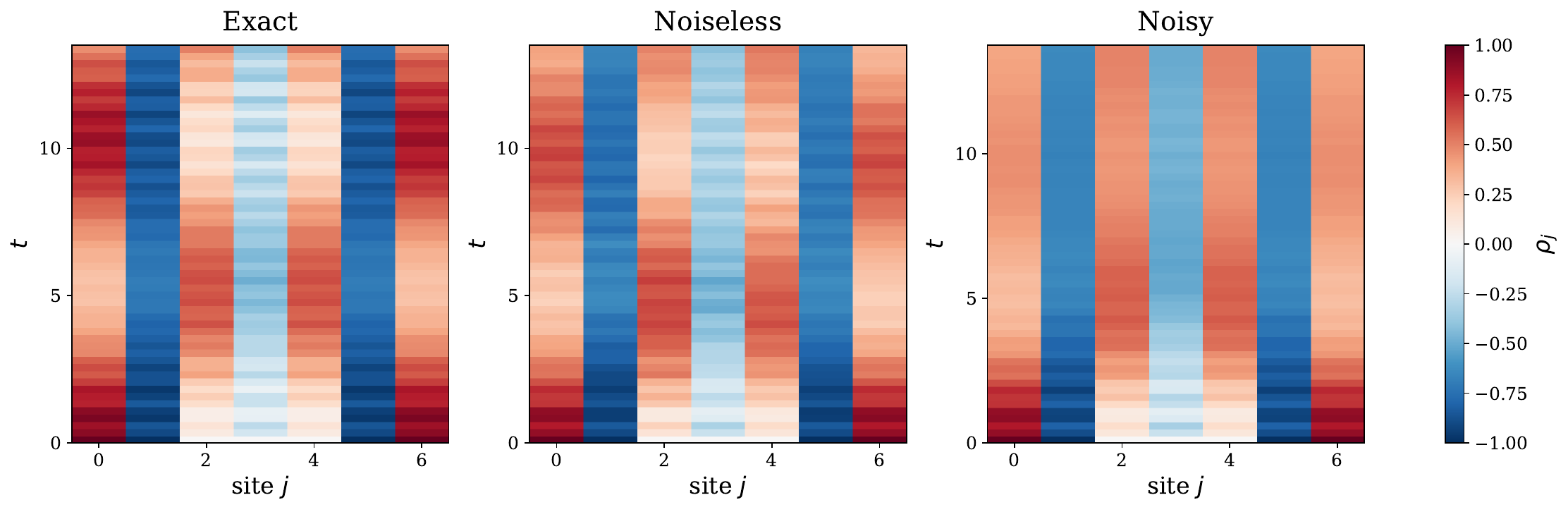}
	\caption{Charge density plots for early times where one sees prominent meson flipping with $\mu=1$, $g=3$, $\kappa =1$ and Trotter step size of $T=0.25$. We compare the exact dynamics (left) with the noisy matter-integrated-out CX gate based circuit (middle) and the noisy matterful MS gate based circuit with post-selection (right).}
	\label{fig:noiseComparison}
\end{figure*}

\subsection{Meson-antimeson collision}
Our study of meson-antimeson collisions consists of examining two distinct parameter regimes in a chain of $11$ matter sites to reveal contrasting scattering behaviors: 
\begin{itemize}
    \item A large-$g$ regime, where the particles reflect off each other, and
    \item A small-$g$ regime, where they pass through one another without interacting.
\end{itemize}
Figure~\ref{meson_antimeson}(a) shows the results for the large-$g$ case, with parameters $\mu = 1$, $g = 3$, and $\kappa = 1$.  As in meson-meson collisions discussed earlier, the meson and antimeson approach each other, interact, and reflect, rather than continuing unimpeded, which we can see from the difference from the free-particle simulations in Fig.~\ref{meson_antimeson_subtracted}(a). The resulting local charge distribution becomes nonzero starting around the time of collision and remains so afterward, confirming that the interaction leads to reflection.

\begin{figure*}
	\centering
	\includegraphics[width=1.02\linewidth]{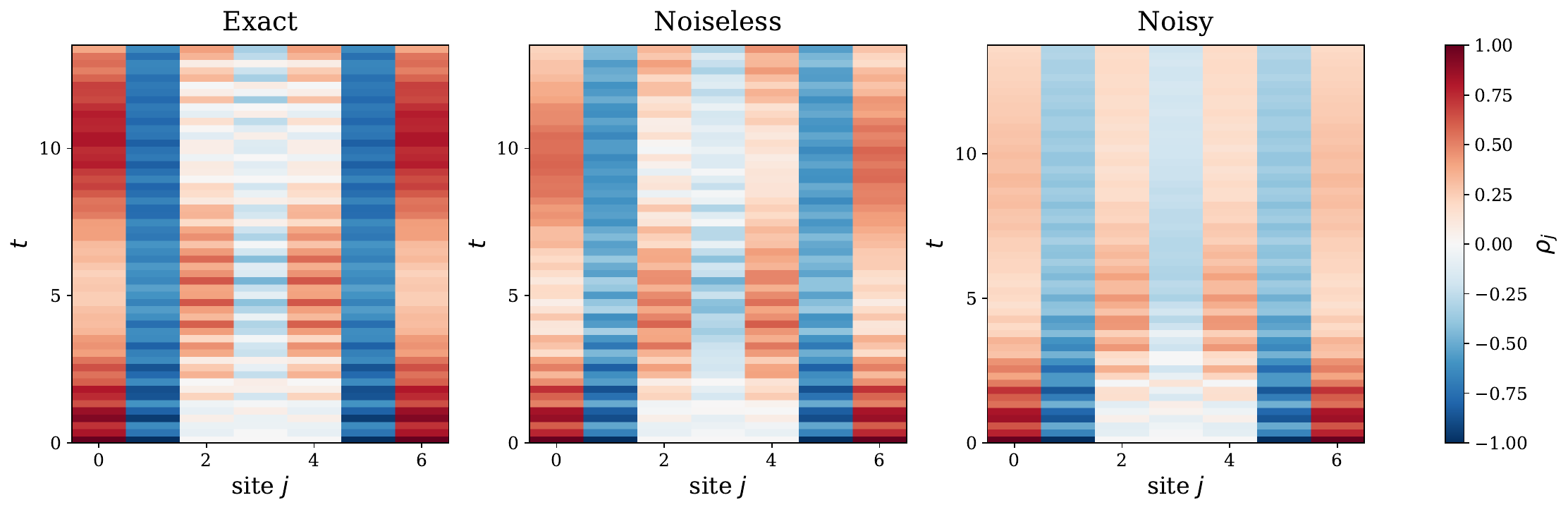}
	\caption{Charge density plots with vacuum fluctuations subtracted off for $\mu=1$, $g=3$, $\kappa =1$ and Trotter step size of $T=0.25$. We compare the exact dynamics (left) with the noisy matter-integrated-out CX gate-based circuit (middle) and the noisy matter-included MS gate-based circuit with post-selection (right).}
	\label{fig:noiseComparison2}
\end{figure*}

\begin{figure}
	\centering
	\includegraphics[width=0.8\linewidth]{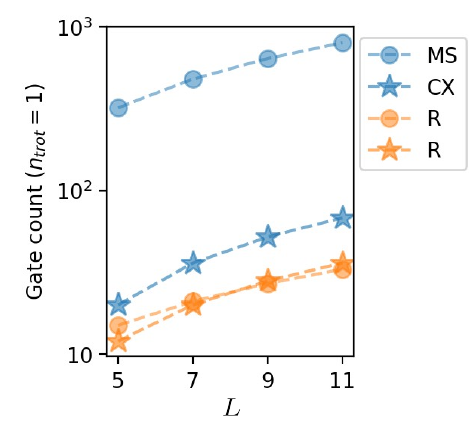}
	\caption{Gate counts over a single Trotter step in the circuit with matter (dashed dotted lines) and for the circuit without matter (dashes starred lines). The two-body gates are the Mølmer--Sørensen (MS) gates and CX gates, while the only single-body gates are the R gates.}
	\label{fig:gatecount} 
\end{figure} 

For the small-$g$ regime, the results are depicted in Fig.~\ref{meson_antimeson}(b). The raw scattering data alone does not clearly indicate whether any interaction takes place, but the difference from the free-particle simulations, plotted in Fig.~\ref{meson_antimeson_subtracted}(b), remains close to zero at all times, even after the collision. This suggests that for small $g$, the meson and antimeson propagate independently and pass through each other without significant interaction. It is natural to interpret this as evidence for the absence of scattering. However, one should note the significant vacuum fluctuations, as seen in Fig.~\ref{meson_antimeson}(b), which means that coherent particle-like behavior gets blurred. This occurs because our initial states consist of bare mesons and antimesons, which are not eigenstates of the free theory and therefore overlap with both the vacuum and excited sectors. 

In contrast, Ref.~\cite{Rigobello2021entanglement} constructs quasiparticle mesons and antimesons as wave packets composed of eigenstates of the free theory ($g = 0$). The more localized these packets are in momentum space, the more closely they approximate true free-theory eigenstates. For such states, the behavior is qualitatively different: even at small $g$, dispersion is minimal and vacuum excitations are strongly suppressed, resulting in much cleaner dynamics. This is further aided by their use of a significantly larger system size ($L = 200$), which allows wave packets to be spread out in real space, thereby narrowing their momentum distribution according to the uncertainty principle. As such, while some dispersion is still expected at $g = 0$ due to the nonlinear dispersion relation and finite width, it remains negligible compared to what we observe in our smaller-scale setup.

Furthermore, in Figs.~\ref{meson-meson_g=3_snapshots}, \ref{meson_antimeson_g=3_snapshots}, and \ref{meson-antimeson_g=0.5_snapshots} in Appendix~\ref{LinkApp}, we plot snapshots of the local expectation value of the electric flux on each link for the noisy, noiseless, and exact cases.
These show more clearly the excellent agreement between the circuit simulations (even with noise) and the exact dynamics.
\section{The system with matter}

Given the above results for the matter-integrated-out model, it is interesting to investigate what happens if the matter fields are kept explicitly in the model.
Such a comparative analysis is crucial, as the process of integrating out matter becomes more challenging --- or even intractable --- in higher dimensions with fermionic matter. 

We can already predict what the contrasting behavior will be by considering the symmetry subsectors in the presence of noise. In an ideal, closed system, the gauge invariance is exactly preserved. However, this is no longer true for real quantum devices subject to noise. This should be contrasted with true gauge invariance in field theories such as QED or QCD, where gauge symmetry is a redundancy in the description of physical states rather than a physical symmetry that can be broken. Physical observables must commute with the symmetry generators, and the physical states must be gauge-invariant. As a result, gauge symmetry cannot be spontaneously broken (Elitzur’s theorem)~\cite{Elitzur1975Dec}.

By contrast, in quantum simulations, gauge symmetry is engineered rather than being a fundamental property of the system, and can thus be explicitly broken by noise. Since all circuits are coupled to an environment, the dynamics is not exactly unitary. As such, noise can drive the system out of its physical subsector, which is only a tiny fraction of the total Hilbert space. Indeed, for $L=7$ we merely have $33$ physical states, compared to a total dimension of $2^7 \times 3^6 \approx 2^{17}$ on the device. For $L = 8$, the full space grows to $\approx 2^{22}$ , while the number of physical states is $61$. In these regimes, leakage into unphysical sectors under noise is highly probable, as the dimension of the physical subspace is a fraction that is smaller than $0.05 \%$ of the total Hilbert space. The situation significantly improves in the matter-integrated-out formulation of the circuits, where the Hilbert space on the device is reduced to $3 \times 4^{L-2}$ states, eliminating the $2^L$ factor from matter degrees of freedom. Although one might initially anticipate a dimension of $3^{L-1}$ states, the actual count of $3 \times 4^{L-2}$ arises from our specific encoding scheme. While unphysical configurations still exist, they now constitute a significantly smaller proportion of the Hilbert space. This is a key reason for the improved noise resilience of the matter-integrated-out circuit observed in Fig.~\ref{fig:noiseComparison}. The Hilbert space leakage into the unphysical subsector is quite severe for the circuit with matter as seen in App.~\ref{ErrorApp}. To mitigate it, we will use post-selection in the circuit with matter, which effectively means that after every Trotter step, we project back onto the physical subsector.

\subsection{The circuit with matter}
\label{circ_with_mat}
We now perform a comparative study to investigate the impact of including matter fields in the circuit. This will enable us to assess the scalability of our approach to higher spatial dimensions, where integrating out the matter is not as feasible. To simulate the dynamics, we perform a second-order Trotter decomposition. The non-interacting contribution \eqref{non-int} can be implemented by a sequence of virtual $RZ$ gates \eqref{virz}. For a single link and the neighboring two matter sites this corresponds to the sequence of gates given in Fig.~\ref{fig:enter-label-2}, where the first and last rows represent qubits and the middle row corresponds to the qutrit.

The minimal coupling term \eqref{minCouple2} contribution can be decomposed into a sequence of three-body gates as 
\begin{align}
	e^{-\imath T \hat{H}_{j}^{\text{min}}} &= e^{-\imath T \kappa (\hat{\sigma}^{+}_{j} \hat{s}^+_{j,j+1} \hat{\sigma}^{+}_{j+1} + \text{H.c.})}\nonumber\\ &= \hat{U}_{j}^{\text{min}}(T\kappa/\sqrt{2}) \hat{P}_{j,+} \hat{U}_{j}^{\text{min}}(T\kappa/\sqrt{2}) \hat{P}_{j,+}^{\dag}, \label{MSdecomp}
\end{align}
where the permutation operator is defined as $\hat{P}_{+;j} = \hat{R}_{Y;j}^{01}(\pi)\hat{R}_{Y;j}^{12}(\pi)$ which cyclically permutes the spin-$1$ states as $\ket{0} \to \ket{1}$, $\ket{1} \to \ket{2}$, and $\ket{2} \to \ket{0}$. Here, the $\hat{R}_{Y;j}^{ab}(\theta)$ gate is a rotation between the qutrit levels $a$ and $b$, defined as
\begin{equation}
    \hat{R}_{Y;j}^{ab}(\theta) \equiv \exp[-\imath \frac{\theta}{2} \hat{\sigma}^{y;ab}_{j,j+1}].
\end{equation}

The minimal coupling term contribution between the states $\ket{0}$ and $\ket{1}$ can be decomposed into $12$ two-body MS  gates~\cite{S_rensen_2000,andrade_engineering_2022}. 
\begin{align}
	\hat{U}_{j}^{\text{min}}(\theta)  \equiv&  \hat{M}_{j;0}^{zy}(-\pi/2)
	\hat{M}_{j;1}^{xx}(\theta)
	\hat{M}_{j;0}^{zy}(\pi/2) \nonumber \\
	&\hat{M}_{j;0}^{yy}(-\pi/2) 
	\hat{M}_{j;1}^{zy}(\theta)
	\hat{M}_{j;0}^{yy}(\pi/2) \nonumber \\
	&\hat{M}_{j;0}^{yx}(-\pi/2) 
	\hat{M}_{j;1}^{zx}(-\theta)
	\hat{M}_{j;0}^{yx}(\pi/2) \nonumber \\
	&\hat{M}_{j;0}^{xx}(-\pi/2) 
	\hat{M}_{j;1}^{zy}(-\theta) \hat{M}_{j;0}^{xx}(\pi/2) \label{UminDecomp}
\end{align}
where $M$ denotes the MS gate defined as
\begin{align}
	\hat{M}_{j;s}^{\mu\nu}(\alpha)  \equiv  e^{\imath \frac{\alpha}{4} (\hat{\sigma}_{j}^{\mu;01} + \hat{\sigma}_{j+s}^\nu)^2}, \label{MSgate}
\end{align}
with $j$ indexing the qutrit and $\hat{\sigma}_{j+s}^{\nu}$ denoting the standard Pauli matrices ($\nu = x, y, z$) acting on the qubit at site $j+s$.
In our circuit implementation, we exclusively use RXX gates, denoted as $\hat{M}_{j;s}^{xx}(\alpha)$. To realize interactions involving different Pauli bases, we perform appropriate basis transformations using single-qudit rotations. For example, 
\begin{subequations}
\begin{align}
    \hat{M}_{j;s}^{zx}(\alpha) &= \hat{R}^{01}_{Y;j}(\pi/2)\hat{M}_{j;s}^{xx}(\alpha)\hat{R}^{01}_{Y;j}(-\pi/2)\\
    \hat{M}_{j;s}^{yx}(\alpha) &= \hat{R}^{01}_{Z;j}(-\pi/2)\hat{M}_{j;s}^{xx}(\alpha)\hat{R}^{01}_{Z;j}(\pi/2).
\end{align}
\end{subequations}

Putting everything together, the dynamics of a single minimal coupling term can be simulated by the circuit given in Fig.~\ref{fig:enter-label2}. 

\section{Noise comparison between matter and without matter}
\label{NoiseComp}

We now compare the results of the circuit with matter and the matter-integrated-out circuit to the exact results by analyzing the local charge densities, shown in Fig.~\ref{fig:noiseComparison}. Notably, when matter is included, no visible scattering is observed, unlike in the other two cases. Instead, the system becomes homogeneous rather quickly, essentially thermalizing due to the noise, after which the dynamics nearly freeze out. To further highlight the differences among the three cases, we also plot the charge densities with the background vacuum fluctuations subtracted off in Fig.~\ref{fig:noiseComparison2}.
While the matter-integrated-out circuit still matches well with the exact case, the matterful circuit becomes locally charged everywhere. For larger noise, the differences between the two circuits becomes even more apparent, as shown in Fig.~\ref{fig:linkexpectations} and discussed in App. \ref{ErrorApp}. 

Ideally, we would consider the same process as in Fig.~\ref{meson_antimeson}, namely switching off the walls. However, due to the large Hilbert space dimension for $L=8$ in the matter-included case, the resulting unitary circuits reach terabyte-scale sizes, making classical numerical simulations infeasible. Consequently, we restrict our study to $L=7$ (with no sites beyond the walls) and compare these results to the matter-integrated-out case.

Let us now try to understand the difference in noise robustness between the two primary circuits under consideration, namely the one with matter and the one with matter integrated out. The most evident distinction arises from comparing the gate counts for these circuits, as shown in Fig.~\ref{fig:gatecount}.  As previously noted, the noise associated with one-body gates is minimal, and since virtual RZ gates are noiseless, their counts are omitted. Consequently, the dominant source of noise originates from the two-body gates - the  MS gates for the matter-included circuit and the CX gates for the matter-integrated-out circuit. Here we note a stark difference between the gate counts. At $L=7$, the matter-included circuit employs roughly ten times more two-body gates than the matter-integrated-out circuit, increasing to about sixteen times at the largest system size considered. This significant difference is the primary reason for the enhanced noise resilience observed in the matter-integrated-out circuit. While improvements can be made through optimized circuit design, the inherent scaling is dictated by the number of qudits in the matter-included case, which is exponentially more than in the matter-integrated-out case.

\section{Experimental platform} \label{Sec:experimental}

We propose the implementation of this model on a trapped ion qudit quantum processor \citep{Ringbauer_2022}, encoding the qudits in the electronic states of $^{40}\mathrm{Ca}^+$ ions as shown in Fig.~\ref{fig:lattice_scheme}(c). In such platforms, quantum circuits are realized digitally by sequences of quantum gate operations, which are realized by coherent laser pulses driving single- and two-qudit operations. Entanglement is mediated by coherently exciting the motion of the ion string, favorably via a mode with nearly homogeneous coupling for all participating qudits. We can realize two schemes on such a quantum processor: (1) the bichromatic Mølmer--Sørensen (MS) scheme \cite{S_rensen_2000} (see \eqref{MSgate}), where two ions are addressed simultaneously and motion is only virtually excited, and (2) an extension of the original Cirac-Zoller scheme \cite{Cirac_1995} by sequences of single-qudit operations, temporarily populating the motional mode. We refer to these gates as controlled rotations ($\hat{\mathrm{C}}_{\text{ROT}}$), which are effectively described by
\begin{equation}
    \hat{C}_{\mathrm{ROT}}(\theta, \phi) = \ketbra{c}{c}\otimes R_{ij}(\theta, \phi) + \sum_{k \ne c } \ketbra{k}{k}\otimes I, \label{CNOT}
\end{equation}
where $R_{ij}(\theta, \phi)$ is a rotation in the two-dimensional subspace spanned by $\ket{i}$ and $\ket{j}$, with the control qudit in state $\ket{c}$; any spectator states $\ket{k,l}$ with $k\neq c$ and $l\neq i,j$ are left unaffected. 

\section{Conclusion and outlook}
\label{SecConOut}
In this work, we have demonstrated the effective simulation of (anti)meson dynamics and scattering processes in lattice gauge theories with higher-level representations of the gauge and electric fields using qudit quantum processors. This facilitates probing collisions not accessible in lattice gauge theories with two-level representations of the fields, thereby motivating the use of qudits for shallower quantum circuit depths to study quench dynamics.

Since our model is in one spatial dimension and fermionic coupling is at most nearest-neighbor in range, we were able to integrate out the matter degrees of freedom through Gauss's law. Since this method requires no qubits to encode matter fields, it significantly reduces both the computational resources and the number of quantum gates needed, enabling quantum simulations over longer timescales and larger system sizes. This would greatly facilitate future quantum simulation experiments employing our proposed qudit digital circuits. We also developed qudit circuits and analyzed quench dynamics on them for the original model (with explicit matter fields), and compared the effect of noise on it relative to the matter-integrated-out case. We found that the latter case suppresses noise much more effectively than the case with explicit matter degrees of freedom.

Beyond serving as a testbed for circuit design and error mitigation, our study demonstrates that meson-meson and meson-antimeson scattering dynamics can act as sensitive probes of gauge-invariant interactions. In particular, the transition from reflection to transmission in meson-antimeson collisions across different coupling regimes, as well as the early-time emergence of meson flipping, reflect the interplay between confinement dynamics and local gauge constraints. These processes remain qualitatively robust under realistic noise, indicating current qudit platforms can provide experimentally accessible signatures of nontrivial gauge-theory scattering.

We have focused on a $1+1$D lattice gauge theory, but it would be interesting to consider higher spatial dimensions, which is the current frontier in the field. Whereas integrating out fermionic matter is possible in $1+1$D, this is not usually possible in higher spatial dimensions.
As our simulations with explicit matter fields show higher sensitivity to noise, it is very important when extending our work to two spatial dimensions to investigate more robust error mitigation schemes that allow accessing sufficiently long evolution times beyond the collision cross-section.
Another interesting future direction would be to extend our approach to higher-spin representations of the gauge field and to non-Abelian gauge groups, in order to probe scattering processes with greater relevance to collider physics.

\footnotesize{\begin{acknowledgments}
    The authors are grateful to Rainer Blatt for stimulating discussions. R.J., J.C.L., J.J.O., and J.C.H.~acknowledge funding by the Max Planck Society, the Deutsche Forschungsgemeinschaft (DFG, German Research Foundation) under Germany’s Excellence Strategy – EXC-2111 – 390814868, and the European Research Council (ERC) under the European Union’s Horizon Europe research and innovation program (Grant Agreement No.~101165667)—ERC Starting Grant QuSiGauge. M.R.~and M.M.~acknowledge funding by the European Union under the Horizon Europe Programme---Grant Agreements 101080086---NeQST and by the European Research Council (ERC, QUDITS, 101039522). Views and opinions expressed are however those of the author(s) only and do not necessarily reflect those of the European Union or the European Research Council Executive Agency. Neither the European Union nor the granting authority can be held responsible for them. We also acknowledge support by the Austrian Science Fund (FWF) through the EU-QUANTERA project TNiSQ (N-6001), by the IQI GmbH, and by the Austrian Federal Ministry of Education, Science and Research via the Austrian Research Promotion Agency (FFG) through the project FO999914030 (MUSIQ) funded by the European Union-NextGenerationEU. This work is part of the Quantum Computing for High-Energy Physics (QC4HEP) working group.
\end{acknowledgments}}
\normalsize

\appendix

\setcounter{figure}{0}
\renewcommand\thefigure{A\arabic{figure}}

\section{Flux snapshots}\label{LinkApp}
\begin{figure*}[t]
    \centering
    	\includegraphics[width=\linewidth]{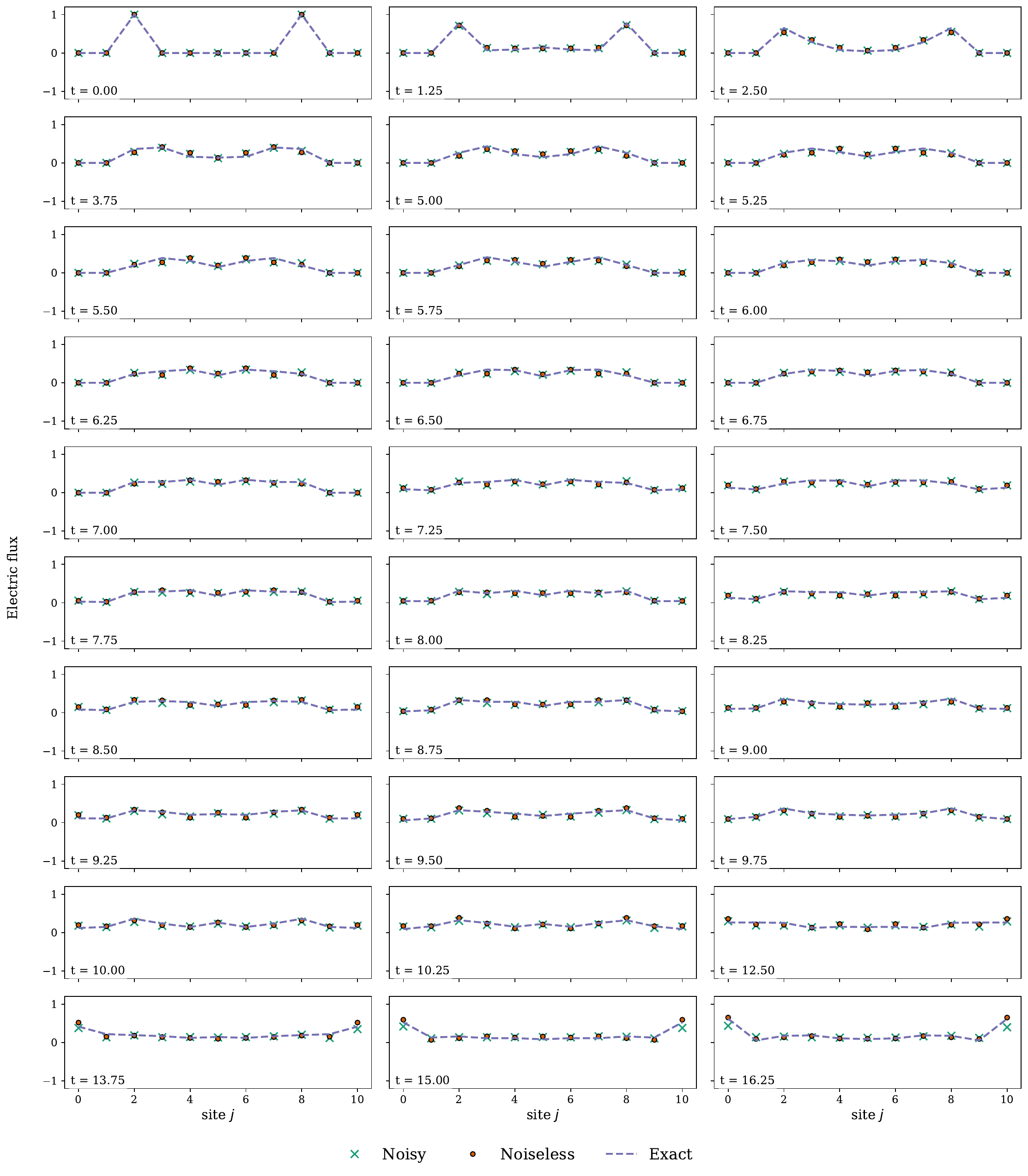}
    \caption{Snapshots for meson-meson collision with the parameters $\mu=1$, $g=3$, $\kappa=1$. The panels display the system at initial times, around the collision time, and at later times when the particles propagate beyond the walls.}
    \label{meson-meson_g=3_snapshots}
\end{figure*}

\begin{figure*}[p]
    \centering
    	\includegraphics[width=\linewidth]{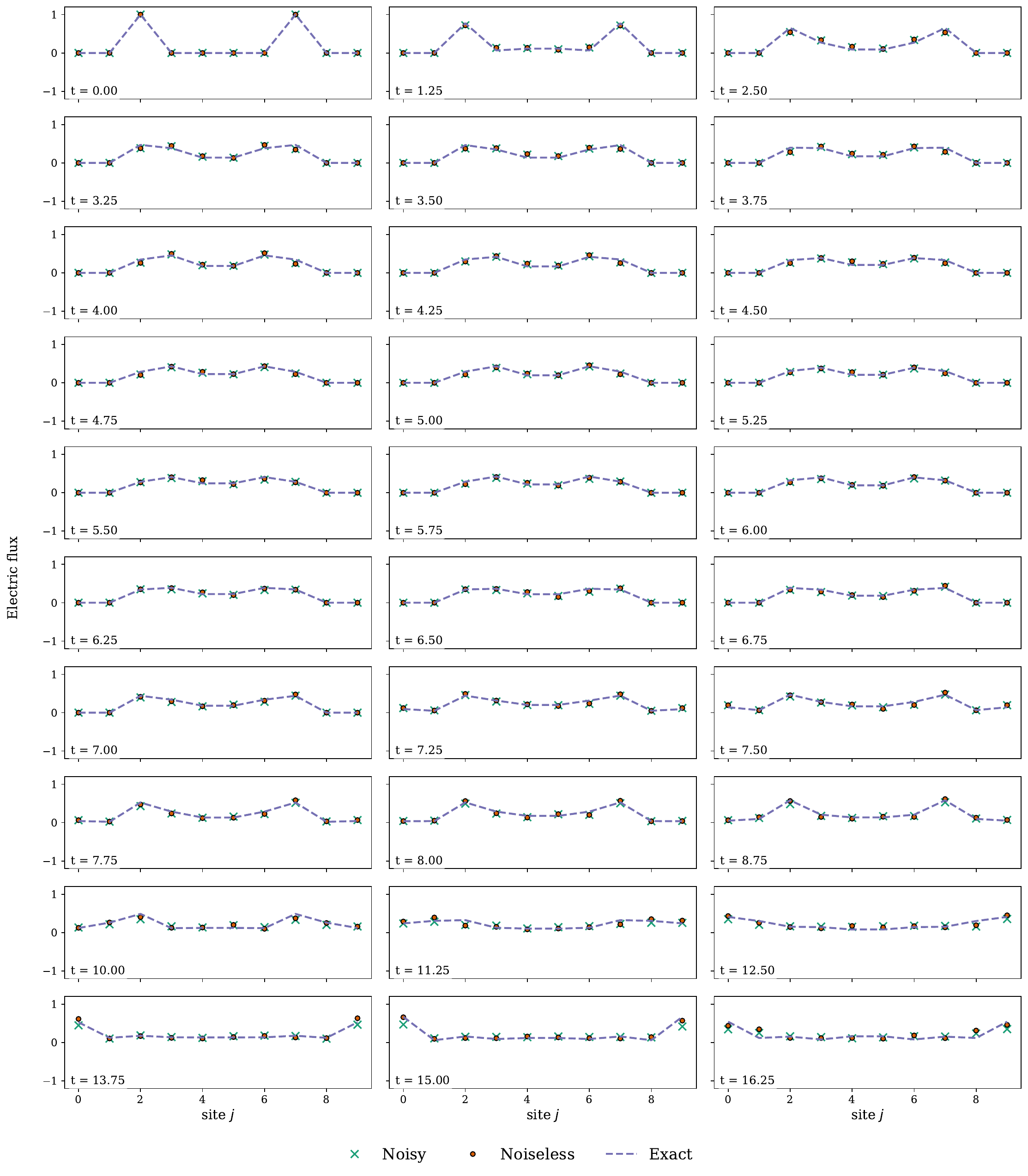}
    \caption{Snapshots depicting the real-time dynamics of a meson-antimeson collision with the parameters $\mu=1$, $g=3$, $\kappa=1$. The panels depict the system before, during, and after the collision, where the meson and antimeson pass through each other.}
    \label{meson_antimeson_g=3_snapshots}
\end{figure*}

\begin{figure*}[p]
    \centering
    	\includegraphics[width=\linewidth]{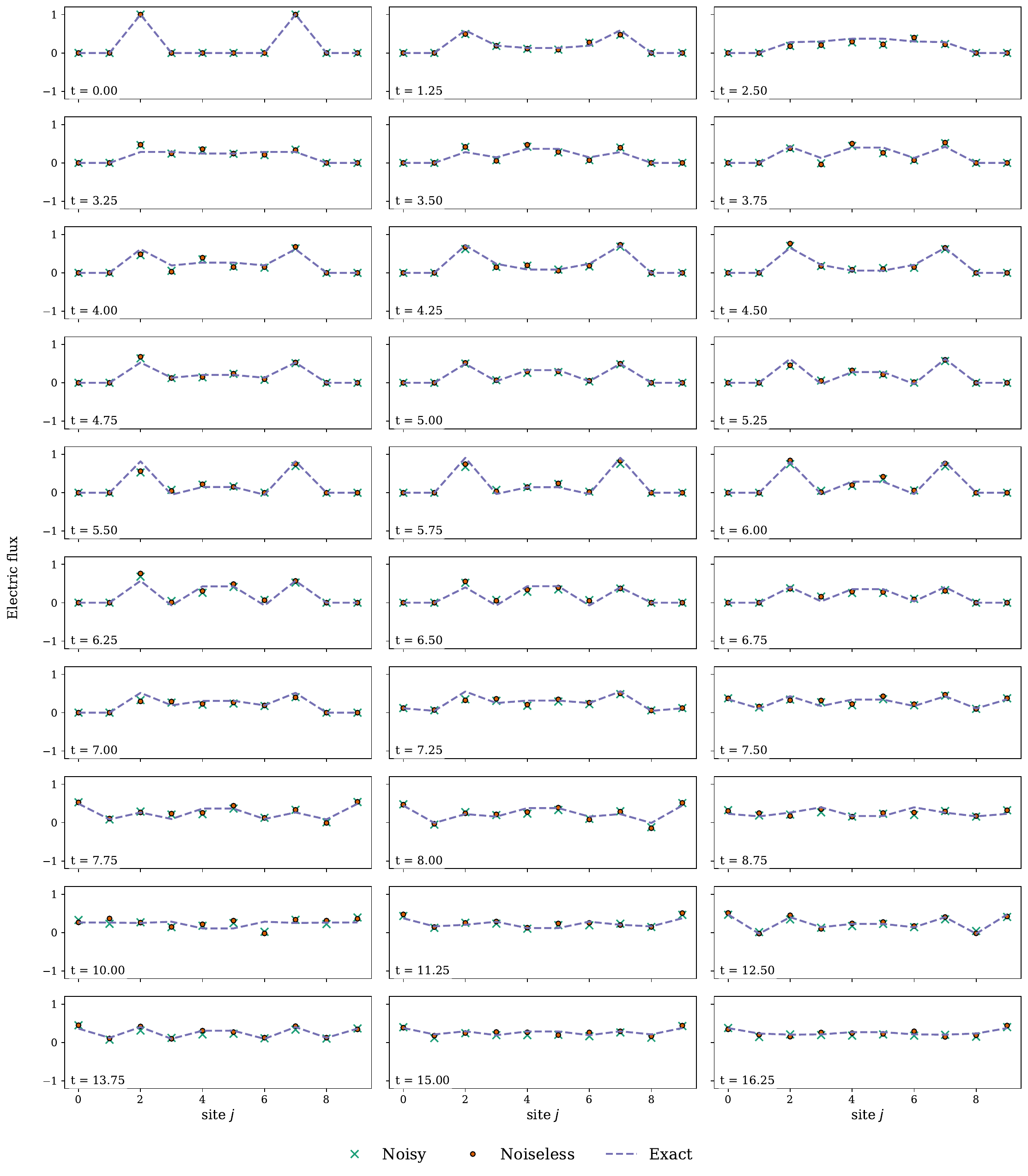}
    \caption{Snapshots for meson-antimeson collision with the parameters $\mu=1$, $g=0.5$, $\kappa=1$. The panels display the system at initial times, around the time of collision, and at later times when the particles have propagated after interacting. }
    \label{meson-antimeson_g=0.5_snapshots}
\end{figure*}

To complement the plots of the evolution of the charge density in the main text, we also present plots for the local electric flux on each link, presented as snapshots for a range of time slices through the evolution.
Figure~\ref{meson-meson_g=3_snapshots} shows snapshots for the meson-meson collision at $g = 3$, while Figs.~\ref{meson_antimeson_g=3_snapshots}, and \ref{meson-antimeson_g=0.5_snapshots} show snapshots for the meson-antimeson collisions at $g = 3$ and $0.5$, respectively.
These plots show more clearly the excellent agreement between the circuit simulations (even with noise) and the exact dynamics.


\newpage

\setcounter{figure}{0}
\renewcommand\thefigure{B\arabic{figure}}

\section{Post-selection and noise}\label{ErrorApp}

\begin{figure}[H]
	\centering
	\includegraphics[width=\linewidth]{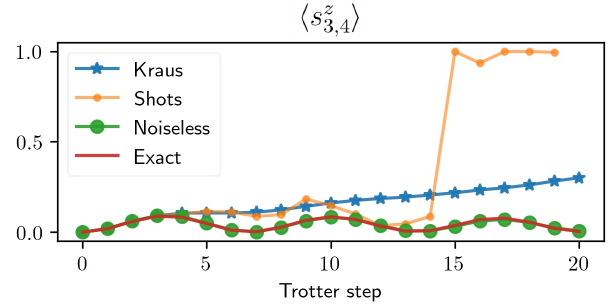}
	\caption{Example of typical discontinuity in shot noise evolution for a local flux. Here we have the Kraus evolution \eqref{KrausEvolution} in blue, the noisy \emph{shots} evolution in orange, the noiseless evolution in green, and the exact time evolution in red.} 
	\label{fig:linkexpectationsJump}
\end{figure}

To see the effect of the Hilbert space leakage in the noisy circuit with matter we study the noisy evolution numerically in two different ways. Both procedures considered are analytically identical but numerically different due to numerical accuracy. The first procedure is the noisy \emph{shots} evolution $\ket{\psi_c(t_{n+1})} \propto \hat{P}_{\text{phys}} \hat{U}_{c}(T) \ket{\psi_c(t_{n})}$
, with the noisy unitary $\hat{U}_{c}(T)$ and physical subspace projector $\hat{P}_{\text{phys}}$. Here the $c$ index indicates that the unitary is noisy and thus different for each shot and each trotter step. We consider a total of $N_s$ shots to average over. Secondly, we consider the Kraus map representation of the noisy state evolution
\begin{equation}
	\hat{\rho}(t_{n+1}) =\sum_{c=1}^{N_s} \hat{K}_{c}(T)  \hat{\rho}(t_n) \hat{K}_{c}(T)^\dag \label{KrausEvolution}
\end{equation}
where the Kraus operators are given as
\begin{equation}
	\hat{K}_{c}(t_{n}) = \hat{P}_{\text{phys.}} \hat{U}_{c}(t_{n})/Z_n
\end{equation}
with normalization constant $Z_n$. 
Note that in both cases we perform the projection onto the physical gauge sector after every Trotter step. This post-selection mitigates the effect of noise and preserves the sparsity of the state (for numerical efficiency). A typical example of how these methods compare is given in Fig. \ref{fig:linkexpectationsJump}.

The noiseless case is very close to the exact case indicating negligible Trotter error. At first both the \emph{noisy shots} and Kraus evolution overlap rather well. Despite the repeated projection back onto the physical subspace, the noisy \emph{shots} evolution has sharp jumps. This is due to normalization issues. Essentially, in some cases most of the state will leak out into the unphysical subsector. The remaining part in the physical subsector is then comparable to numerical error. We still keep this part though, but now it also contains amplified numerical error after we project and renormalize, hence the jumps in Fig.~\ref{fig:linkexpectationsJump}. Such a renormalization need not occur in the Kraus evolution case; thus it lacks these jumps. As such, for the matterful case, we always use the Kraus evolution.

\begin{figure}[H]
	\centering
	\includegraphics[width=0.9\linewidth]{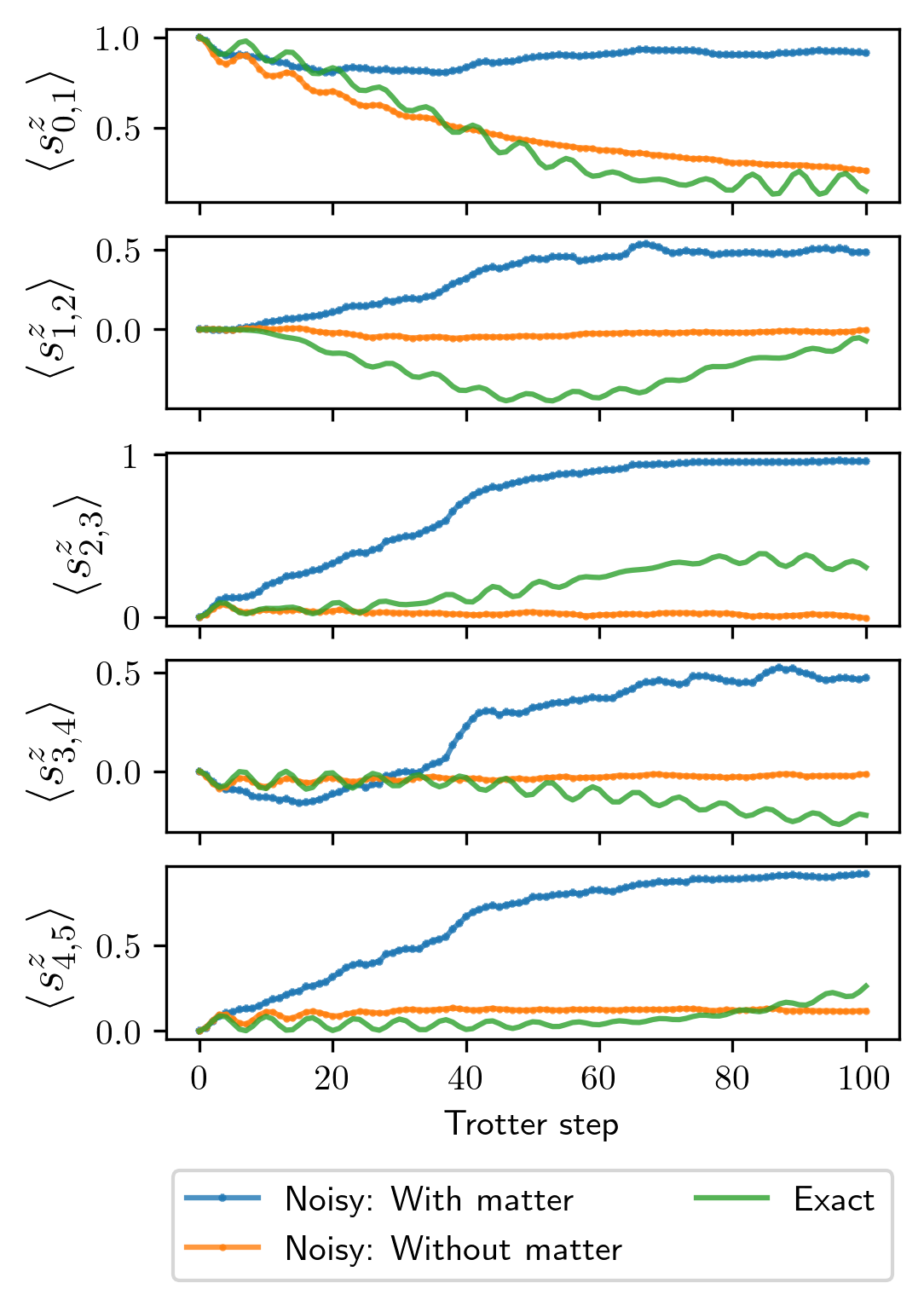}
		\caption{Link fluxes evolved over time assuming noise probabilities $p_{\text{two-body}} \approx 10^{-3}$. The matter-integrated-out results are in orange, while the setup with matter is in blue.} 
	\label{fig:linkexpectations}
\end{figure}

Let us now compare with the case where matter is integrated out (Fig.~\ref{fig:linkexpectations}). It can be observed that there exists a bias for the matter-included circuit in the link values drifting to one extreme. Considering that the main noise stems from the $2$-body gates, which only ever involve the states $\ket{0}$ and $\ket{1}$ in the circuit implementation, one might not be surprised in seeing such a bias. 

One consideration is whether the error is intrinsic and thus unavoidable; if so, the focus should shift to understanding the fundamental limitations and exploring possible mitigation strategies.  The first path stems from the observation that our two-body gate of choice---namely, the standard MS gate---does not commute with the gauge generators $\hat{G}_j$, although the full sequence of gates $\hat{U}(T)$ does (this can be seen in how the interaction terms in \eqref{MSdecomp} commute with $\hat{G}_j$). One might assume that adding noise to such a gate could lead to high error stemming from the breaking of gauge invariance. An alternative is the RZZ gate $\hat{R}_{ZZ;j;s}^{01}(\alpha) \equiv e^{\imath \frac{\alpha}{2} \hat{\sigma}_{j}^{z;01} \hat{\sigma}_{j+s}^z}$.
Theoretically this is essentially the same gate, but with a bases change from $x$ to $z$ bases. However, now the two body gates commute with $\hat{G}_j$. Unfortunately, this yields only a slight error decrease.

\bibliography{biblio}
\end{document}